\documentclass[aps, prc, twocolumn, superscriptaddress, showpacs, floatfix, longbibliography, american]{revtex4-2}

\usepackage{amsfonts,amssymb,latexsym,xspace,epsfig,graphicx,color}
\usepackage{amsmath,enumerate,stmaryrd,xy,stackrel,ulem}
\usepackage{bm}
\usepackage{siunitx}
\usepackage{braket}
\usepackage[utf8]{inputenc}
\usepackage[english]{babel}
\usepackage{xcolor}
\definecolor{tolred}{HTML}{CC3311}
\definecolor{tolmagenta}{HTML}{EE3377}
\definecolor{tolgreen}{HTML}{228833}
\usepackage[colorlinks=true]{hyperref}
\hypersetup{
  linkcolor=tolred,
  citecolor=tolgreen,
  urlcolor=tolmagenta}
\usepackage{microtype}
\usepackage{enumerate}
\usepackage{ifthen}

\newcommand{\nm}{\ensuremath{N_{\mathrm{max}}}}
\newcommand{\w}{\ensuremath{\hbar\omega}}
\newcommand{\YN}{\ensuremath{Y\!N}}
\newcommand{\YNN}{\ensuremath{Y\!N\!N}}
\newcommand{\NN}{\ensuremath{N\!N}}
\newcommand{\NNN}{\ensuremath{N\!N\!N}}
\newcommand{\SNN}{\ensuremath{\Sigma NN}}
\def\nuc#1#2{\relax\ifmmode{}^{#1}{\protect\mathrm{#2}}\else${}^{#1}$#2\fi}
\def\hnuc#1#2{\relax\ifmmode{}_{\Lambda}^{#1}{\protect\mathrm{#2}}\else${}_\Lambda^{#1}$#2\fi}
\newcommand{\nnlosim}{NNLO\textsubscript{sim}}
\newcommand{\nnlosimstd}{\nnlosim(\LamNN = \qty{500}{MeV}, \Tmax =
  \qty{290}{\MeV})}
\newcommand{\loyn}{LO \YN}
\newcommand{\loynstd}{LO {\YN} (\LamYN = \qty{600}{MeV})}
\newcommand{\LamNN}{\ensuremath{\Lambda_{\NN}}}
\newcommand{\LamYN}{\ensuremath{\Lambda_{\YN}}}
\newcommand{\Tmax}{\ensuremath{T_{\mathrm{Lab}}^{\mathrm{max}}}}
\newcommand{\intmsr}[2]{\ensuremath{\mathrm{d} p^{#2}_{#1}\, p^{#2 2}_{#1}}}
\newcommand{\tbd}{\ensuremath{\hyp \to \he+\pi^-}}
\newcommand{\Gtbd}{\ensuremath{\Gamma(\tbd)}}
\newcommand{\luv}{\ensuremath{\Lambda_{\mathrm{UV}}}}
\newcommand{\lir}{\ensuremath{L_{\mathrm{IR}}}}
\newcommand{\hyp}{\hnuc{3}{H}}
\newcommand{\he}{\nuc{3}{He}}

\newcommand{\nmmax}{68}
\newcommand{\BRval}{0.35(4)}
\newcommand{\BL}{\ensuremath{B_\Lambda}}
\newcommand{\tl}{\ensuremath{\tau_\Lambda}}
\newcommand{\thyp}{\ensuremath{\tau(\hyp)}}
\newcommand{\pim}{\ensuremath{\pi^-}}
\newcommand{\piz}{\ensuremath{\pi^0}}

\newcommand{\cgC}[6]{C_{#1,#2;#3,#4}^{#5;#6}}

\newcommand{\DeltaI}[0]{\ensuremath{\Delta T}}

\newcommand{\newabbreviation}[3]{\newcounter{#1}\expandafter\newcommand\csname#1\endcsname[1][]{\ifthenelse{\equal{##1}{abreviate}}{#2}{\ifthenelse{\equal{##1}{fullname}}{#3}{\ifthenelse{\equal{##1}{explain}}{#3
          (#2)\stepcounter{#1}}{\ifthenelse{\value{#1}=0}{#3##1
            (#2##1)\stepcounter{#1}}{#2##1}}}}}}
\newabbreviation{NCSM}{\text{NCSM}}{no-core shell model}
\newabbreviation{LO}{\text{LO}}{leading order}
\newabbreviation{IR}{\text{IR}}{infrared}
\newabbreviation{UV}{\text{UV}}{ultraviolet}
\newabbreviation{HO}{\text{HO}}{harmonic oscillator}
\newabbreviation{gs}{\text{g.s.}}{ground-state}
\newabbreviation{CG}{CG}{Clebsh--Gordan}
\newabbreviation{FSI}{\text{FSI}}{final-state interaction}
\newabbreviation{eft}{EFT}{effective field theory}
\newabbreviation{chieft}{\(\chi\)EFT}{chiral effective field theory}
\newabbreviation{RHI}{RHI}{relativistic heavy-ion}
\newabbreviation{DW}{DW}{distorted wave}
\newabbreviation{PW}{PW}{plane wave}
\newabbreviation{CM}{CM}{center of mass}
\newabbreviation{PC}{PC}{parity-conserving}
\newabbreviation{PV}{PV}{parity-violating}
\newabbreviation{BC}{BC}{bubble-chamber}

\begin{document}

\title{Hypertriton lifetime}

\author{D.~Gazda}
\thanks{corresponding author, gazda@ujf.cas.cz}
\affiliation{Nuclear Physics Institute, 25068 \v{R}e\v{z}, Czech Republic}

\author{A.~P\'{e}rez-Obiol}
\affiliation{Barcelona Supercomputing Center, 08034 Barcelona, Spain}

\author{E.~Friedman}
\affiliation{Racah Institute of Physics, The Hebrew University, Jerusalem 9190401, Israel}

\author{A.~Gal}
\affiliation{Racah Institute of Physics, The Hebrew University, Jerusalem 9190401, Israel}

\date{\today}

\begin{abstract}
  Over the last decade, conflicting values of the hypertriton {\hyp}
  lifetime {\thyp} were extracted from {\RHI} collision experiments,
  ranging from values compatible with the free-\(\Lambda\) lifetime
  {\tl}---as expected naively for a very weakly bound \(\Lambda\) in
  {\hyp}---to lifetimes as short as \(\thyp \approx (0.4-0.7)\,\tl\).
  In a recent work~\cite{Perez-Obiol:2020qjy} we studied this {\hyp}
  lifetime puzzle theoretically using realistic three-body {\hyp} and
  {\he} wave functions computed within the ab~initio no-core shell
  model approach with interactions derived from chiral effective field
  theory to calculate the partial decay rate \Gtbd{}. Significant but
  opposing contributions were found from \SNN{} admixtures in
  \hyp{} and from \(\pim{}-\he\) final-state interaction. In
  particular, {\thyp} was found to be strongly correlated with the
  \(\Lambda\) separation energy {\BL} in {\hyp}, the value of which is
  rather poorly known experimentally and, in addition, is known to
  suffer from sizable theoretical uncertainties inherent in the
  employed nuclear and hypernuclear interaction models. In the present
  work we find that these uncertainties propagate into {\thyp}, and
  thus limit considerably the theoretical precision of its computed
  value. Although none of the conflicting {\RHI} measured {\thyp}
  values can be excluded, but rather can be attributed to a poor
  knowledge of {\BL}, we note the good agreement between the lifetime
  value \(\thyp=\qty[parse-numbers=false]{242(28)}{\ps}\) computed
  at the lowest value \(\BL=\qty{66}{\keV}\) reached by us and the
  very recent ALICE measured lifetime value
  \(\tau^{\text{ALICE}}(\hyp) =
  \qty[parse-numbers=false]{253(11)(6)}{\ps}\) associated with the
  ALICE measured {\BL} value
  \(B^{\text{ALICE}}_{\Lambda} =
  \qty[parse-numbers=false]{102(63)(67)}{\keV}\)~\cite{ALICE:2022sco}.
\end{abstract}

\maketitle

\section{Introduction}
\label{sec:introduction}

\noindent The hypertriton ({\hyp}) is the lightest bound hypernucleus,
with isospin \(T=0\) and spin-parity
\(J^P=\frac12^+\)~\cite{Gal:2016boi}. Together with other \(s\)-shell
light $\Lambda$ hypernuclei, \hnuc{4}{H}--\hnuc{4}{He} and
\hnuc{5}{He}, it provides useful constraints on the poorly known \YN{}
and \YNN{} interactions~\cite{Le:2023bfj} which are tested in heavier
$\Lambda$ hypernuclei in the context of hyperon composition of dense
baryonic matter realized perhaps in the cores of neutron
stars~\cite{Logoteta:2019utx,Gerstung:2020ktv,Friedman:2022bpw2023ucs}.
Owing to the extremely small \(\Lambda\) separation energy in {\hyp},
\(\BL=\qty{164(43)}{\keV}\)~\cite{Hypweb}, the structure of {\hyp}
should resemble to a good approximation that of a \(\Lambda\) bound
loosely to a deuteron (\nuc2H) core with a mean distance of
\(\approx\qty{10}{\femto\meter}\). The lifetime of such a loosely
bound system is expected to be comparable to the lifetime of the free
\(\Lambda\),
\(\tl=\qty[parse-numbers=false]{263(2)}{\ps}\)~\cite{ParticleDataGroup:2020ssz},
the decay of which to \qty{99.7}{\percent} is driven by the
\(\Lambda\to N\,\pi\) nonleptonic weak transitions.

\begin{figure}[tb]
  \centering
  \includegraphics[width=\columnwidth]{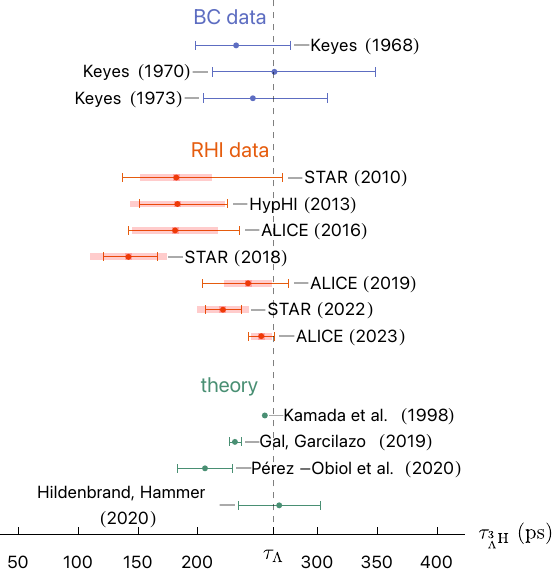}
  \caption{{\hyp} lifetime values {\thyp} obtained in
    selected~\cite{Eckert:2022dyz} past nuclear
    \BC{}~\cite{PhysRevLett.20.819,KEYES1970,Keyes:1973two} and recent
    {\RHI}
    collision~\cite{STAR:2010gyg,Rappold:2013fic,ALICE:2015oer,STAR:2017gxa,ALICE:2019vlx,STAR:2021orx,ALICE:2022sco}
    experiments, together with a few (post-1997) microscopic
    calculations~\cite{Kamada:1997rv,Gal:2018bvq,Perez-Obiol:2020qjy,Hildenbrand:2020kzu};
    see text. Error bars and shaded areas indicate the measurement
    statistical and systematical errors, respectively, as well as the
    estimated theoretical uncertainties (if available). The vertical
    dashed line corresponds to the free-\(\Lambda\) lifetime value
    \(\tl=\qty[parse-numbers=false]{263(2)}{\ps}\)~\cite{ParticleDataGroup:2020ssz}.
  }
\label{fig:exp_th_data}
\end{figure}

Yet, while the most recent ALICE Collaboration's
measurement~\cite{ALICE:2022sco} reports a \thyp{} closely agreeing
with the free-\(\Lambda\) lifetime \tl{},
\(\tau^{\text{ALICE}}(\hyp) =
\qty[parse-numbers=false]{253(11)(6)}{\ps} \approx \qty{0.96}{\tl}\),
considerably shorter values
\(\thyp\approx\qty[parse-numbers=false]{(0.5-0.7)}{\tl}\) were
extracted in recent {\RHI} collision experiments by the
STAR~\cite{STAR:2017gxa} and HypHI~\cite{Rappold:2013fic}
Collaborations. We note that the latest STAR Collaboration's
measurement~\cite{STAR:2021orx}, reporting a value of
\(\tau^{\text{STAR}}(\hyp) =
\qty[parse-numbers=false]{221(15)(19)}{\ps}\), comes close within its
experimental uncertainties to {\tl}, but its central value is still
about \qty{15}{\percent} shorter. It is to be noted that a similarly
large spread of {\thyp} values, although with larger uncertainties,
had been reported in older nuclear-emulsion and helium \BC{}
hypernuclear measurements~\cite{Lock:1964bp,PhysRev.180.1307,
  PhysRevLett.20.819,KEYES1970,Bohm:1970se,Keyes:1973two}, as shown in
Fig.~\ref{fig:exp_th_data}.
Also shown in the figure are the {\thyp} values from recent {\RHI}
collision
experiments~\cite{STAR:2010gyg,Rappold:2013fic,ALICE:2015oer,STAR:2017gxa,ALICE:2019vlx,STAR:2021orx,ALICE:2022sco},
together with representative (post-1997) theoretical
calculations~\cite{Kamada:1997rv,Gal:2018bvq,Perez-Obiol:2020qjy,Hildenbrand:2020kzu}.

Several theoretical approaches with varying degree of sophistication
have been employed to study the decay of the
hypertriton~\cite{Rayet:1966fe,Congleton:1992kk,Kamada:1997rv,Gal:2018bvq,Perez-Obiol:2020qjy,Hildenbrand:2020kzu}
by calculating {\pim} decay rates and using the \(\DeltaI = \frac12\)
rule (see Sec.~\ref{sec:lifetime}) to add the corresponding {\piz}
decay rates. In the first calculation marked in
Fig.~\ref{fig:exp_th_data}, a value of {\thyp} shorter by only
\qty{6}{\percent} than {\tl} was obtained in a full \(\Lambda n p\)
Faddeev calculation accounting for all {\hyp} two-{}, three-{} and four-body
{\pim} decay channels (\(\nuc{3}{He}\,\pim\),
\(\nuc{2}{H}\, p\,\pim\), and \(n\,p\,p\,\pim\), all with plane-wave
pions). However, the Nijmegen SC89 {\YN}
interaction~\cite{Maessen:1989sx} applied there to construct the
{\hyp} wave function was shown to be problematic in hypernuclei,
starting at \(A=4\)~\cite{Nogga:2001ef}. In the second
calculation~\cite{Gal:2018bvq}, a value of
\(\thyp\approx \qty{0.8}{\tl}\) was obtained within the closure
approximation using three-body {\hyp} wave functions generated from
\(\Lambda n p\) Faddeev equations. Here, half of the
\(\approx\qty{20}{\percent}\) reduction in {\thyp} resulted from the
attractive {\FSI} of the outgoing pion. While a fixed value of
\(\BL \approx \qty{135}{\keV}\) was used in these two {\thyp}
calculations, a range of {\BL} values was tested in the next two
calculations. The {\thyp} intervals shown in the figure for these
calculations~\cite{Perez-Obiol:2020qjy,Hildenbrand:2020kzu} correspond
to the central value \(\BL=\qty{102}{\keV}\) reported by the most
recent ALICE measurement~\cite{ALICE:2022sco}. In the third
calculation~\cite{Perez-Obiol:2020qjy} we established significant but
opposing contributions from pionic {\FSI} and \SNN{} admixtures
in {\hyp}, by using {\chieft} nuclear and hypernuclear interactions to
obtain realistic {\he} and {\hyp} three-body wave functions within an
ab~initio {\NCSM} approach for evaluating the two-body {\tbd} partial
decay rate. The known branching ratio
\({\Gtbd} / \Gamma(\hyp\to\text{all~}\pim\text{~channels})\) was then
utilized to compute the inclusive {\pim} decay rate. Finally,
\(\thyp\approx\tl\) was obtained in the last calculation displayed in
the figure~\cite{Hildenbrand:2020kzu}, with relatively large
uncertainties of \(\gtrsim \qty{15}{\percent}\), by applying pionless
{\eft} and reducing the three-body {\hyp} and \nuc{3}{H}, \nuc{3}{He}
systems to loosely bound \(\nuc{2}{H}{-}\Lambda\) and
\(\nuc{2}{H}{-}N\) two-body systems, respectively. Adding pionic
{\FSI} was found in a very recent application of this pionless {\eft}
work to enhance the {\hyp} decay rate by about
\qty{18}{\percent}~\cite{Hildenbrand:2023rkc}. The central value of
the {\thyp} interval in Fig.~\ref{fig:exp_th_data} moves then from
about {\tl} down to \qty{84}{\percent} of it, namely to
\qty{221}{\ps}, in rough agreement with our
result~\cite{Perez-Obiol:2020qjy}.

In the present work, we extend and provide full details of our recent
\hyp{} lifetime calculation~\cite{Perez-Obiol:2020qjy}. More
specifically, we employ precise realistic three-body {\hyp} and {\he}
wave functions to evaluate the {\hyp} two-body \pim{} partial decay
rate and relate it to the inclusive pionic {\hyp} decay rate and
{\thyp}. We account for the distortion of the outgoing \pim{} wave due
to realistic \({\pim}-{\he}\) interaction and take into account
contributions from \(\Sigma\to N\,\pi\) transitions, resulting from
\SNN{} admixtures in the {\hyp} wave function. The main focus of the
present work, however, is to explore the theoretical precision of
calculating {\Gtbd}, and thereby {\thyp}, due to systematic model
uncertainties in the {\YN} and {\NN}+{\NNN} interactions. For this
purpose we employ four versions of the {\loyn} interaction regularized
at \(\LamYN = \qtylist[parse-numbers=false,
list-units=single]{550;600;650\text{,};700}{\MeV}\) together with the
{\nnlosim} family of 42 nuclear interactions.

The paper is organized as follows: In Sec.~\ref{sec:formalism} we
present the formalism to calculate the hypertriton decay rate and
lifetime. In particular, in Sec.~\ref{sec:lifetime} we elaborate on
our approach to relate \thyp{} to the partial two-body \pim{} decay
rate \Gtbd{} introduced in Sec.~\ref{sec:decayrate}. This partial
decay rate is calculated starting from an effective weak-decay
operator constructed in Sec.~\ref{sec:operator} and evaluating its
matrix elements between \pim{} (Sec.~\ref{sec:pionwf}), \hyp{} and
\he{} three-body wave functions (Sec.~\ref{sec:nuclearwf})
generated within the NCSM approach using realistic \YN{} and
\NN+\NNN{} interactions from \chieft{} described in
Sec.~\ref{sec:int}. In Sec.~\ref{sec:results}, results for the
two-body \tbd{} decay rate are presented in Sec.~\ref{sec:tbdr},
identifying its dominant contributions in
Sec.~\ref{sec:decayrateapprox} and studying in
Sec.~\ref{sec:luvtuning} its dependence on the {\hyp} \(\Lambda\)
separation energy {\BL}. In Sec.~\ref{sec:unc} we quantify the
precision limits in predicting theoretically \Gtbd{}, arising from the
uncertainties in the nuclear and hypernuclear three-body wave
functions. In Sec.~\ref{sec:res:lifetime}, we present and discuss
results for the hypertriton lifetime, particularly in light of recent
\thyp{} measurements. Our main findings are summarized in
Sec.~\ref{sec:conclusions}.

\section{Method}
\label{sec:formalism}

\subsection{Hypertriton decay}
\label{sec:lifetime}

\noindent The main decay channels which contribute to the total decay
rate rate of the hypertriton are the mesonic (nonleptonic) decay modes
due to the weak-interaction \(\Lambda \to N\,\pi\) transitions:
\begin{equation}
  \label{eq:mesonic}
  \begin{aligned}
    \hyp &\to \he+\pi^-,            &\hyp &\to \nuc{3}{H}+\pi^0, \\
    \hyp &\to \nuc{2}{H} +p+\pi^-,  &\hyp &\to \nuc{2}{H} + n+\pi^0, \\
    \hyp &\to p+p+n+\pi^-,          &\hyp &\to n+n+p+\pi^0.
  \end{aligned}
\end{equation}
This is in contrast with heavier hypernuclei where the mesonic decays
are Pauli blocked. In~\eqref{eq:mesonic}, only the four-body
deuteron-breakup channels are heavily suppressed due to the limited
phase space~\cite{Kamada:1997rv}. Apart from the mesonic decays, there
is also the nonmesonic decay branch of \hyp{} due to
\(\Lambda N \to NN\):
\begin{equation}
  \begin{aligned}
    \hyp &\to \nuc{2}{H} + n, \\
    \hyp &\to  n+n+p.
  \end{aligned}
\end{equation}
The nonmesonic modes, while being the major contributors in the decays
of heavy hypernuclei, are known to play a very small role in
\hyp{}~\cite{Rayet:1966fe,Golak:1996hj,Perez-Obiol:2018oax}.

The {\hyp} partial decay rates corresponding to the charged and
neutral pion channels listed in~\eqref{eq:mesonic} are not
independent. Since for the experimental ratio
\(\Gamma(\Lambda\to n\,\pi^0)/\Gamma(\Lambda\to p\,\pi^-)\approx 0.5\)
holds to a good precision, the isospin \(T=0 \to \frac32\)
(\(\DeltaI = \frac32\)) components of the
\(\Lambda\to p\,\pim, n\,\pi^0\) amplitudes must be negligible in
comparison with the \(\DeltaI=\frac12\) amplitudes. This is known as
the ``\(\DeltaI = \frac12\) rule'' which implies that the transition
operators satisfy
\(\hat{O}_{\Lambda\to p\pi^-}=\sqrt{2}\,\hat{O}_{\Lambda\to n\pi^0}\)
and thus relates the charged and neutral pionic rates by
\(\Gamma_{\pi^-}=2\,\Gamma_{\pi^0}\). In addition, the measured
world-average \BC{} branching ratio
\(R_3\equiv\Gtbd/\Gamma_{\pi^-}(\hyp)=\BRval\)~\cite{Keyes:1973two}
can be used to relate the inclusive {\pim} rate
\(\Gamma_{\pim}(\hyp)\) with the two-body \pim{} rate
\(\Gamma^{\he}=\Gtbd\).

To summarize, in this work we employ the following strategy to
evaluate the \hyp{} lifetime: (i) Use the branching ratio \(R_3\) to
get the inclusive {\pim} rate from the two-body {\pim} rate \Gtbd{} as
\begin{equation}
  \begin{split}
    \Gamma_{\pi^-}(\hyp) &= \Gtbd\\
    &+ \Gamma(\hyp\to \nuc{2}{H}+p+\pi^-)\\
    &+ \Gamma(\hyp\to n+p+p+\pi^-)\\
    &= \frac{1}{R_3}\Gtbd{}.
  \end{split}
\end{equation}
(ii) Include the \(\pi^0\) decay channels by employing the empirical
\(\DeltaI = \frac12\) rule
\begin{equation}
  \Gamma_{\pi}= \Gamma_{\pi^-}+ \Gamma_{\pi^0} = \frac32\, \Gamma_{\pi^-}= \tau_{\pi}^{-1}(\hyp).
\end{equation}
(iii) Account for the nonmesonic \(\Lambda N\to NN\) and pionic
true-absorption \(\pi + NN\to NN\) contributions through an increase
of the \hyp{} decay rate by \qty{1.5}{\percent} and
\qty{0.8}{\percent}~\cite{Rayet:1966fe,Golak:1996hj,Perez-Obiol:2018oax},
respectively; \(\thyp{} = 0.978 \,\tau_{\pi}(\hyp)\).

\subsection{Two-body decay rate \texorpdfstring{\Gtbd{}}{Gamma(L3H to
    pi minus + 3He)}}
\label{sec:decayrate}

\noindent We follow Ref.~\cite{Kamada:1997rv} which relates the \hyp{}
two-body \pim{} decay rate \Gtbd{} in the total momentum zero frame to
the \(\Lambda\to N\,\pim\) free-\(\Lambda\) weak-decay vertex operator
\(\hat{O}\) as
\begin{equation}
  \label{eq:decayrate1}
  \begin{split}
    \Gamma^{\he} &= \frac12\sum_{m_{\hyp}}\sum_{m_{\he}} \int
                   \mathrm{d}^3p_{\he}\frac{\mathrm{d}^3p_\pi}{8\pi^2 E_\pi}\\
                 &\times \left\lvert \sqrt{3}\, \langle
                   \Psi_{\he;\vec{p}_{\he},m_{\he}}\phi_{\pi;\vec{p}_\pi} \vert  \hat{O}\vert\Psi_{\hyp;m_{\hyp}}\rangle\right\rvert^2\\
                 &\times \delta^{(3)}(\vec{p}_{\he}+\vec{p}_\pi) \\
                 &\times \delta{\left(M_{\hyp}-M_{\he}-\frac{\vec{p}_{\he}^{\;2}}{2M_{\he}}
                   -E_\pi\right)}.
  \end{split}
\end{equation}
The expression involves averaging over the initial, \(m_{\hyp}\), and
summation over the final, \(m_{\he}\), spin projections of the
(hyper)nuclear states \(\Psi\) and the integration over the two final
\he{} and \pim{} momenta, \(\vec{p}_{\he}\) and \(\vec{p}_\pi\), is
accompanied by the corresponding phase factor, with
\(E_\pi=\sqrt{m_\pi^2+\vec{p}_\pi^2}\) the relativistic energy of the
pion. The final \(\pim - \he\) scattering state
\(\ket{\Psi_{\he}\phi_{\pi}}\) is discussed in detail in
Sec.~\ref{sec:pionwf}. The isospin factor \(\sqrt{3}\) in
Eq.~\eqref{eq:decayrate1} accounts for the three final nucleons into
which the \(\Lambda\) may transition and the Dirac delta functions
ensure momentum and energy conservation by fixing the values of
\(\vec{p}_{\he}\) and the modulus of the outgoing pion momentum
\(\vec{p}_\pi\), where \(M_{\hyp}=\qty{2991}{\MeV}\) and
\(M_{\he}=\qty{2809}{\MeV}\) in the argument are the \hyp{} and \he{}
masses. Since there is no preferred spatial direction, we choose the
quantization \(z\) axis in the direction of \(\vec{p}_\pi\) and
integrate out the remaining angular dependence as
\(\int \mathrm{d}\hat{p}_\pi=4\pi\). Eq.~\eqref{eq:decayrate1} then
becomes
\begin{equation}
  \label{eq:decayrate2}
  \begin{split}
    \Gamma^{\he} &= \frac{3}{4\pi}
                   \frac{M_{\he}\,q_\pi}{M_{\he}+E_\pi} \\
                 &\times \sum_{m_{\hyp}}\sum_{m_{\he}} \left\lvert
                   \braket{\Psi_{\he}\phi_\pi|\hat{O}|\Psi_{\hyp}}
                   \right\rvert^2,
  \end{split}
\end{equation}
where the pion momentum is kinematically fixed by
\begin{equation}
  \label{qpimod}
  \begin{split}
    q_\pi &= \sqrt{2M_{\he}}\\
          &\times \sqrt{M_{\hyp}
            -\sqrt{m_\pi^2+2M_{\he}M_{\hyp}-M_{\he}^2}} \\
          &= \qty{114.4}{\MeV}
  \end{split}
\end{equation}
for average pion mass \(m_\pi=\qty{138.04}{\MeV}\) and corresponds to
pion energy \(E_\pi=\sqrt{m_\pi^2+q_\pi^2}=\qty{179.3}{\MeV}\).

\subsection{Weak-decay operator}
\label{sec:operator}

\noindent The transition operator in Eq.~\eqref{eq:decayrate2}
originates from an effective Lagrangian density for the weak
\(\Lambda\to N\,\pi\) transitions
\begin{equation}
  \label{eq:phenlagrangian}
  \mathcal{L}_{\Lambda N\pi} =
  G_Fm_\pi^2\,\bar{\psi}_N(\mathcal{A}_\Lambda + \mathcal{B}_\Lambda
  \gamma_5) (\vec{\tau}\cdot\vec{\pi})\,\psi_{\Lambda},
\end{equation}
where \(G_Fm_{\pi}^2=2.21 \times 10^{-7}\); \(\vec{\tau}\) are the
isospin Pauli matrices; and \(\psi_{\Lambda}\), \(\psi_N\), and
\(\vec{\pi}\) are the \(\Lambda\), nucleon, and isovector pion fields,
respectively. The \(\mathcal{A}_\Lambda=1.024\) and
\(\mathcal{B}_\Lambda=-9.431\) are the {\PV} spin-dependent and {\PC}
spin-independent \(\Lambda\to N\,\pi\) amplitudes. Values of these
amplitudes were fixed by the lifetime of free \(\Lambda\) and the
PC/PV decay rates ratio,
\(\Gamma_{\mathrm{PC}} / \Gamma_{\mathrm{PV}} = \num{0.203}\),
determined from the BESIII value of the \(\Lambda\to p\,\pi^-\)
asymmetry parameter~\cite{BESIII:2018cnd}. The empirical
\(\DeltaI = \frac12\) rule, discussed in Sec.~\ref{sec:lifetime}, is
incorporated formally in Eq.~\eqref{eq:phenlagrangian} by adding
\(\vec{\tau}\) at the \(\Lambda N \pi\) vertex, introducing a spurious
\(\ket{\frac12,-\frac12}\) isospin to the \(\Lambda\) hyperon, and
assuming isospin conservation~\cite{Kamada:1997rv}.

In its nonrelativistic form, the transition operator
\begin{equation}
  \label{eq:weakopl}
  \hat{O} = i
  \sqrt{2}G_Fm_\pi^2\left(
    {\cal A}_\Lambda+
    \frac{{\cal B}_\Lambda}{2\overline{M}_{\Lambda N}}
    \vec{\sigma}\cdot \vec{q}_\pi\
  \right)\hat{P}^{(\Lambda)}_{t_{12}=0},
\end{equation}
derived from the Lagrangian density~\eqref{eq:phenlagrangian}, acts on
the \(A=3\) wave functions and is responsible for the transition of
the \(\Lambda\) hyperon in {\hyp} to a proton in {\he}. The factor
\(\sqrt{2}\) is due to the spurious isospin of the \(\Lambda\), and the
projection operator \(\hat{P}^{(\Lambda)}_{t_{12}=0}\) selects the
isospin \(t_{12}=0, t_3 = 0\) \(\Lambda NN\) channels in the \hyp{}
wave functions. The \(\vec{\sigma}\) are the Pauli matrices in spin
space, \(\vec{q}_\pi=q_\pi\hat{z}\) is the pion (on-shell) momentum,
and \(\overline{M}_{\Lambda N}=\frac{1}{2}(M_\Lambda+M_N)\),
calculated using the average nucleon mass \(M_N=\qty{938.92}{\MeV}\)
and \(M_\Lambda=\qty{1115.68}{\MeV}\).

Considering the \SNN{} admixtures in \hyp{} induced by the
\(\Lambda N\leftrightarrow \Sigma N\) coupling in the hypernuclear
Hamiltonian, new \(\Sigma^-\to n\,\pi^-\) and
\(\Sigma^0 \to p\,\pi^-\) transitions become available and contribute
to \Gtbd{}. To account for the \(\Sigma\to N\,\pi\) contributions, we
generalize the weak-decay operator as
\begin{equation}
  \label{eq:weakop}
  \hat{O}\to \hat{O}
  +i\,G_Fm_\pi^2\left(
    \frac{\sqrt{2}}{3}{\cal A}_{\Sigma^-} + \frac{1}{3}{\cal A}_{\Sigma^0}\right)\hat{P}^{(\Sigma)}_{t_{12}=1}.
\end{equation}
The factors \(\frac{\sqrt{2}}{3}\) and \(\frac{1}{3}\) arise due to
isospin \CG{} coefficients when embedding the \(\Sigma^-\) and
\(\Sigma^0\) amplitudes within our \(A=3\) isospin states and
\(\hat{P}^{(\Sigma)}_{t_{12}=1}\) selects the isospin
\(t_{12} =1 , t_3 = 1\) \SNN{} channels in the {\hyp} wave
functions. In Eq.~\eqref{eq:weakop}, we neglect the PC part of
\(\Sigma\to N\,\pi\) amplitudes and fix the \(\Sigma^-\to n\,\pi^-\)
PV amplitude, \(\mathcal{A}_{\Sigma^-}=1.364\), to satisfy the
\(\Sigma^-\) weak-decay lifetime value
\(\tau_{\Sigma^-}=\qty{147.9}{\ps}\)~\cite{Donoghue:1992dd}. In the
case of \(\Sigma^0\), the main decay mode is the electromagnetic
\(\Sigma^0\to \Lambda\,\gamma\) transition, rendering the mesonic
modes negligible. For the \(\Sigma^0\to p\,\pi^-\) amplitude we thus
use the chiral-Lagrangian prediction to relate
\(\mathcal{A}_{\Sigma^0}=\frac{1}{\sqrt
  2}\,\mathcal{A}_{\Sigma^-}\)~\cite{Donoghue:1992dd}.

The explicit form of the transition operator~\eqref{eq:weakop} matrix
elements between \hyp{} and \(\he{}-\pim\) wave functions in
Eq.~\eqref{eq:decayrate2} can be found in
Appendix~\ref{sec:matrixelements}.

\subsection{Pion wave function}
\label{sec:pionwf}

\noindent The {\DW} pion wave function
\(\phi_{\pi}(\vec{q}_\pi;\vec{r}) \equiv
\braket{\vec{r}\,|\phi_{\pi;\vec{q_\pi}}}\) input to the transition
matrix element in Eq.~\eqref{eq:decayrate1} was generated from a
standard optical
potential~\cite{Friedman:2007zza,Friedman:2014msa,Friedman:2019zhc}.
It evolves via {\FSI} from a {\PW} pion with momentum \({\vec q}\) in
the \hnuc{3}{H} rest frame, and its argument
\({\vec r}=\frac{2}{3}{\vec r_3}\) is identified with the coordinate
of the third ``active'' baryon with respect to the {\CM} of \nuc{3}{He}.

As a first approximation, the pion wave function is considered to be a
{\PW} with on-shell momentum \(\vec{q}_\pi\) in the \(z\) direction,
\(\vec{q}_\pi = q_\pi\hat{z}\). Accordingly, a state
\(\ket{q_\pi\hat{z}}\) is inserted in the matrix element and fixes the
momentum difference, \(\vec{p}_3^{\,\prime}-\vec{p}_3\), between the
two active baryons. The next step is to generate pion radial functions
in the presence of a realistic pion--nucleus optical model. As in
our previous publication~\cite{Perez-Obiol:2020qjy}, a natural
starting point is provided by optical potentials that reproduce the
vast amount of experimental strong interaction level shifts and widths
in pionic atoms, essentially at zero energy.

Pion wave functions distorted by the pion-nucleus interaction, are
obtained by calculating pion-nucleus bound states in standard complex
optical potentials. In our previous
publication~\cite{Perez-Obiol:2020qjy} we used global fits to pionic
atoms derived from least-squares fits to 100 pionic atom data from
\nuc{}{Ne} to \nuc{}{U}. In the present work we have extended the data
base by including eight additional species below \nuc{}{Ne}, including
also \nuc{3}{He}. The new potential parameters~\cite{Fri20} are
consistent with those used in our previous publication and the added
\(\chi^2\) for \num{16} points was \num{14.5}. For \nuc{3}{He} we
found \(\chi^2=\num{1.8}\) for two points. Moving over to the true
energy of the pion in the present process is relatively simple as
described below.

The commonly accepted pion-nucleus optical potential is made of an
\(s\)-wave term and a \(p\)-wave term, each containing a real part
linear in the nuclear density and a complex part quadratic in density,
representing pion absorption on two nucleons. The real coefficients
turned out to be close to the corresponding spin-{} and
isospin-dependent amplitudes for the free pion-nucleon interaction at
threshold whereas the complex coefficients were phenomenological, with
poorly determined real parts. Consequently we could set these real
parts to zero without affecting the quality of fits to experiment.
Then the real coefficients could be identified with the corresponding
free pion-nucleon amplitudes at threshold. To extrapolate from
near threshold to \(q_\pi=\qty{114.4}{\MeV}\) in the
\(\pi^--\nuc{3}{He}\) {\CM} system we revised the above \(\pi\,N\)
linear-density terms using energy-dependent scattering amplitudes from
the SAID package~\cite{Arndt:2006bf}. For the nonlinear terms, we
extrapolated their threshold values using as an additional point fits
to \(\pi^{\pm}\) elastic scattering at
\(T_{\rm Lab}=\qty{21.5}{\MeV}\) on \nuc{}{Si}, \nuc{}{Ca}, \nuc{}{Ni},
and \nuc{}{Zr}~\cite{Friedman:2004jh,Friedman:2005pt}. This resulted
in a practically vanishing value of the \(s\)-wave quadratic term and
a \qty{65}{\percent} increase of the \(p\)-wave term.

Expanding \(\phi_{\pi}({\vec{q}_\pi};{\vec r})\) in our calculations in partial
waves \(l_{\pi}\), and recalling the spin-parity \(J^P={\frac{1}{2}}^+\)
of both {\he} and {\hyp}, it follows that the only values allowed are
\(l_{\pi}=0,2\).
Numerically we find a negligible \(l_{\pi}=2\) contribution
of order \qty{0.1}{\percent}, proceeding exclusively through the relatively
minor {\PC} amplitude which in total contributes \(\lesssim \qty{3}{\percent}\) to
{\Gtbd}.

To insert the {\DW}, derived in position space, in the transition
matrix element, we first Fourier transform it to momentum space using
\begin{equation}
  \tilde{\phi}_{l_\pi}(k) = Y_{l_\pi
    0}^{\,*}(\hat{z})\frac{2}{\pi\,q_\pi} \int_0^{\infty} \mathrm{d}r\, r j_{l_\pi}(k r)\phi_{l_\pi}(q_\pi;r)
\end{equation}
and express the {\pim} state \(\ket{\phi_\pi}\) in
Eq.~\eqref{eq:decayrate1} in terms of partial-wave basis states
\(\ket{k\,l_\pi}\) as
\begin{equation}
  \label{eq:phipi}
  \ket{\phi_\pi} = \sum_{l_\pi}\int \mathrm{d}k\,k^2\,\tilde{\phi}_{l_\pi}(k)\ket{k\,l_\pi}.
\end{equation}
The state is normalized as
\begin{equation}
  \label{eq:normphipi}
  \sum_{l_\pi}\int \mathrm{d} k\, k^2 \lvert\tilde{\phi}_{l_\pi }(k)\rvert^2=1.
\end{equation}
All details of the derivation and numerical implementation are given
in Appendix~\ref{app:ft}.

\subsection{Nuclear and hypernuclear wave functions}
\label{sec:nuclearwf}

\noindent The initial-{} and final-state wave functions of {\hyp} and
{\he} in Eq.~\eqref{eq:decayrate1} were computed within the ab~initio
no-core shell model (NCSM) \cite{Barrett2013, Wirth:2017bpw}. In this
approach, nuclei and hypernuclei are described as systems of \(A\)
nonrelativistic point-like particles interacting through realistic
nucleon--nucleon (\NN), three-nucleon (\NNN), and hyperon-nucleon (\YN)
interactions. In NCSM, the many-body wave function is expanded in a
complete set of \HO{} basis states characterized by the \HO{}
frequency \(\w\) and truncated by the maximum number \nm{} of \HO{}
excitations above the lowest configuration allowed by Pauli principle,
\begin{equation}
  \label{eq:expansion}
  \ket{\Psi^{J^\pi T}} = \sum_{N=0}^{\nm} \sum_\lambda c_{N\lambda}^{J^\pi T}
  \ket{N\lambda JT}.
\end{equation}
Here, \(N\) is the total number of \HO{} excitations of all particles and
\(J^{\pi}T\) are the total angular momentum, parity, and isospin. The
quantum number \(\lambda\) labels all additional quantum numbers and the sum
over \(N\) is restricted by parity to an even or odd sequence. The
energy eigenstates are obtained by solving the  Schr\"odinger equation.

In this work we employed a version of NCSM formulated in
translationally invariant relative Jacobi-coordinate \HO{} basis, which
is suitable for dealing with few-body systems. Different sets of
Jacobi coordinates can be employed, one of which is particularly
convenient for the construction of \HO{} basis states antisymmetric
with respect to all nucleons and evaluation of the matrix element in
\eqref{eq:decayrate1}. For \(A=3\) (hyper)nuclear systems, we
introduce
\begin{equation}
  \label{eq:jac}
  \begin{split}
    \vec{p}_{\mathrm{CM}} &= \vec{p}_1+\vec{p}_2+\vec{p}_3,\\
    \vec{p}_{12} &= \frac{1}{m_1 + m_2} \left( m_2\, \vec{p}_1 - m_1\, \vec{p}_2\right),\\
    \vec{p}_3 &= \frac{1}{\sum_{i=1}^3m_i} \left[
      m_3(\vec{p}_1+\vec{p}_2) - (m_1+m_2)\,\vec{p}_3\right],
  \end{split}
\end{equation}
where \(m_i\) and \(\vec{p}_i\) are the mass and momentum of particle
\(i = 1, 2, 3\). In \eqref{eq:jac}, \(\vec{p}_{\mathrm{CM}}\) is the
\CM{} momentum, \(\vec{p}_{12}\) is the relative momentum of two
nucleons, and \(\vec{p}_3\) is the momentum of the third particle
(nucleon or hyperon) with respect to the CM of the nucleon pair.

The use of relative coordinates allows us to separate out and omit the
CM degrees of freedom. Consequently, the \HO{} basis states in
\eqref{eq:expansion} for \hyp{} can be constructed as
\begin{equation}
  \label{eq:expansion3LH}
  \ket{N \lambda J T}_{\hyp} = \ket{(\lambda_{12},\lambda_3)J T},
\end{equation}
where
\(\ket{\lambda_{12}} \equiv \ket{n_{12}(l_{12}s_{12})j_{12}t_{12}}\)
are two-nucleon \HO{} states, depending on the coordinate
\(\vec{p}_{12}\), with radial \(n_{12}\), orbital \(l_{12}\), spin
\(s_{12}\), angular momentum \(j_{12}\), and isospin \(t_{12}\)
quantum numbers. Antisymmetry of the \NN{} states with respect to the
nucleon interchange is achieved by imposing
\((-1)^{l_{12}+s_{12}+t_{12}}=-1\). Similarly, the \HO{} states
\(\ket{\lambda_3} \equiv \ket{n_3(l_3s_3)j_3t_3}\) depending on the
coordinate \(\vec{p}_3\) describe the relative motion of the hyperon
(\(\Lambda\) or \(\Sigma\) for \(t_3 = 0, 1\)) with respect to the CM
of the \NN{} pair. The number of \HO{} excitations in the state
\eqref{eq:expansion3LH} is \(N=2\,n_{12}+l_{12}+2\,n_3+l_3\) and the
parentheses denote angular momentum and isospin coupling. In the case
of \he{}, the \HO{} states in \eqref{eq:expansion} have to be
antisymmetric with respect to exchanges of all nucleons. The
antisymmetrization procedure, when relative Jacobi coordinates are
employed, is discussed in detail, e.g., in
Ref.~\cite{Navratil:1999pw}. The fully antisymmetric \HO{} states are
obtained as linear combinations of \HO{} states with a lower degree of
antisymmetry,
\begin{equation}
  \label{eq:expansion3He}
    \ket{N \lambda J T}_{\he} = \sum_{\tilde{\lambda}}
    C^{NJT}_{\lambda\tilde{\lambda}} \ket{(\tilde{\lambda}_{12},\tilde{\lambda}_3)JT},
\end{equation}
where the expansion coefficients \(C^{NJT}_{\lambda \tilde{\lambda}}\)
are the coefficients of fractional parentage. Here, analogously to
Eq.~\eqref{eq:expansion3LH}, \(\ket{\tilde{\lambda}_{12}}\) depending
on \(\vec{p}_{12}\) is an antisymmetric \HO{} state describing the
relative motion of the \NN{} pair, and \(\ket{\tilde{\lambda}_3}\)
depending on \(\vec{p}_3\) is associated with the relative motion of
the third nucleon with respect to the CM of the \NN{} pair.

To evaluate the matrix elements of the weak-decay operator between the
initial and final states, we project the \hyp{} and \he{} NCSM wave
functions \eqref{eq:expansion} onto a momentum-space partial-wave
basis as
\begin{equation}
  \label{eq:wfpw}
  \begin{split}
    \ket{\Psi^{J^\pi T}} &= \sum_{\alpha}\int
                           \intmsr{12}{}\,\intmsr{3}{}\,
                           \psi_{\alpha}(p_{12},p_3)\\
                         &\times \ket{p_{12}\,p_3\,\alpha}.
  \end{split}
\end{equation}
Here, \(p_{12(3)} = \lvert \vec{p}_{12(3)}\rvert \), and \(\alpha\)
labels the \(JT\)-coupled three-particle channels,
\begin{equation}
  \label{eq:alpha}
  \Ket{\alpha} \equiv \ket{\left((l_{12}s_{12})j_{12}(l_3s_3)j_3\right)JM}
  \Ket{(t_{12}t_3)TM_T},
\end{equation}
where \(M\) and \(M_T\) are the projections of the total angular
momentum and isospin. The labeling of momenta and two-{} and
one-particle spin, isospin, and angular momentum quantum numbers
follows the same scheme as in the case of \HO{} basis states in
Eqs.~\eqref{eq:expansion3LH} and \eqref{eq:expansion3He}. The
coefficient functions \(\psi_\alpha\) in Eq.~\eqref{eq:wfpw} are
combinations of the expansion coefficients in \eqref{eq:expansion},
\eqref{eq:expansion3He}, and momentum-space radial \HO{} functions.
They are normalized as
\begin{equation}
  \label{eq:norm}
  \sum_{\alpha}\int \intmsr{12}{}\, \intmsr{3}{}
  \vert\psi_\alpha(p_{12},p_3)\rvert^2 = 1.
\end{equation}

\subsection{Input nuclear and hypernuclear interactions}
\label{sec:int}

\begin{figure}[t]
  \centering
  \includegraphics[width=\columnwidth]{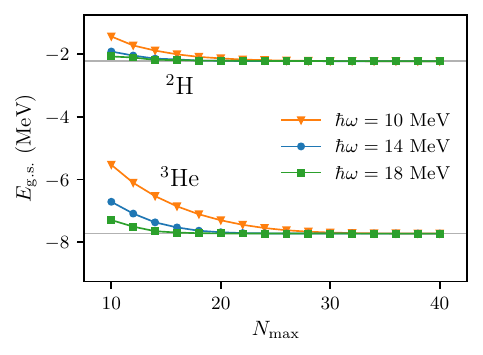}
  \caption{Dependence of \nuc{2}{H} and \he{} \gs{} energies
    \(E_{\mathrm{g.s.}}\) on the NCSM model-space truncation
    \(10 \leq \nm \leq 40\) for several values of the \HO{} frequency,
    \(\hbar\omega = \qtylist[parse-numbers=false,
    list-units=single]{10;14\text{,};18}{\MeV}\), calculated using the
    \nnlosimstd{} interaction. Indicated by the gray horizontal lines
    are the corresponding experimental values.}
  \label{fig:2h3he}
\end{figure}

\noindent In this work, we utilized realistic \NN+\NNN{} and \YN{}
interactions derived from \chieft{}. We did not apply any
renormalization, such as the similarity renormalization group or
Lee--Suzuki transformation, to either of the interactions. In
particular, we employed the \nnlosim{} nuclear forces constructed at
next-to-next-to-leading order (NNLO)~\cite{Carlsson:2015vda} which
were optimized to simultaneously reproduce \NN{} as well as \(\pi N\)
scattering data, the binding energies and charge radii of \nuc{2,3}{H}
and \nuc{3}{He}, the quadrupole moment of \nuc{2}{H}, and the
\(\beta\)-decay half-life of \nuc{3}{H}. The \nnlosim{} is a family of
42 different interactions where each potential is associated with one
of seven regulator cutoffs,
\(\LamNN = 450, 475, \ldots, 575, \qty{600}{MeV}\), together with six
different maximum scattering energies in the laboratory frame,
\(\Tmax = 125, 158, \ldots, 257, \qty{290}{MeV}\), truncating the
experimental \NN{} cross sections used to constrain the respective
interaction. We note that the 42 parametrizations of the nuclear force
give equally good descriptions of the calibration data. Using all of
them allows us to expose the magnitude of systematic nuclear-model
uncertainties resulting from the incomplete knowledge of the nuclear
interaction. In this work we are mainly interested in the wave
function of \he{} and the ``core nucleus'' \nuc{2}{H}. Since certain
low-energy properties of \nuc{2}{H} and \nuc{3}{He} were included in
the pool of fit data, their energies are accurately described for all
these interactions:
\(E_{\nuc{2}{H}, \nuc{3}{He}} =
\qty[parse-numbers=false]{-2.224^{(+0)}_{(-1)},
  -7.717_{(-21)}^{(+17)}}{\MeV}\)~\cite{Carlsson:2015vda}. For
\nnlosim{}, the NCSM-calculated \gs{} energies of \nuc{2}{H} and
\nuc{3}{He} exhibit a weak dependence on the model-space truncation
\nm{} and the \HO{} frequency \w{}, as shown in Fig.~\ref{fig:2h3he}
for \NN+\NNN{} interaction with \(\LamNN = \qty{500}{MeV}\) and
\(\Tmax = \qty{290}{MeV}\). The g.s.\ energies are converged within
few~\qty{}{\keV} already at \(\nm \approx 30\) for a wide range of
\HO{} frequencies \w.

\begin{figure}[t]
  \centering
  \includegraphics[width=\columnwidth]{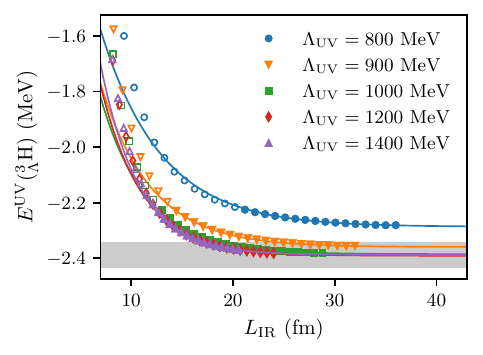}
  \caption{NCSM \hyp{} \gs{} energies \(E^{\mathrm{UV}}({\hyp})\)
    (empty and filled symbols) as functions of the \IR{} length \lir{}
    (up to \(\nm=\nmmax\)) calculated using the \nnlosimstd{} and
    \loynstd{} interactions for several fixed values of the \UV{}
    cutoff \luv{}, together with their extrapolations (solid lines).
    Only the points marked by filled symbols, corresponding to
    particle-stable \hyp{} configurations,
    \(E(\hyp) < E({\nuc{2}{H}})\), are included in the fits. The
    world-average value of the measured {\hyp} energy
    \(E^{\mathrm{exp.}}(\hyp) =
    \qty[parse-numbers=false]{-2.389(43)}{\MeV}\)~\cite{Hypweb} is
    indicated by the gray band.}
  \label{fig:l3h}
\end{figure}

For the \YN{} interaction we employed the Bonn--J\"{u}lich
coupled-channel flavor-SU(3)-based \chieft{}
model~\cite{Polinder:2006zh}. This potential is constructed at {\LO}
and regularized by a smooth momentum cutoff \LamYN{} ranging from
550 to \qty{700}{\MeV}. Its parameters were determined from fits to
the measured \YN{} scattering cross sections, additionally constrained
by allowing for a bound \(J^\pi= \frac{1}{2}^+\) \hnuc{3}{H} state.
This interaction was found to be consistent with the experimental
value of the \(\Lambda\) separation energy in \hyp{}, employing the
NCSM~\cite{Wirth:2017bpw} as well as Faddeev~\cite{Haidenbauer:2007ra}
approaches. NCSM calculations of \hnuc{3}{H} g.s.\ energy exhibit a
stronger, undesired dependence on the model-space truncation \nm{} and
the \HO{} frequency \w{}~\cite{Htun:2021jnu,Gazda:2022fte}. The slow
convergence can be attributed to the small \(\Lambda\) separation
energy and, accordingly, the long tail of the \hyp{} wave function in
position space. The truncation of the \HO{} basis in terms of \nm{}
and \w{} can be translated into associated \IR{} and \UV{} scales. For
NCSM, the corresponding \IR{} and \UV{} scales, \lir{} and \luv{}, can
be extracted by studying the kinetic energy spectrum and used to
extrapolate results obtained in truncated model spaces to infinite
model space (\(\lir \to \infty\))~\cite{Wendt2015,Gazda:2022fte}. The
LO correction for energies and the expected magnitude of subleading
corrections \(\sigma_{\mathrm{IR}}\)~\cite{Forssen:2017wei} are
\begin{equation}
  \label{eq:ir}
  \begin{split}
    E(\lir) &= E_\infty + a_0\, e^{-2k_\infty\lir},\\
    \sigma_{\mathrm{IR}} &\propto \frac{e^{-2k_\infty\lir}}{k_\infty\lir}.
  \end{split}
\end{equation}
The extrapolated energy \(E_\infty\), together with \(a_0\) and
\(k_\infty\), are parameters determined from fits to the
NCSM-calculated energies with weights proportional to the inverse of
\(\sigma_{\mathrm{IR}}\). Here we apply an iterative procedure with
the weights fixed in each optimization until self-consistency for
\(k_\infty\) in Eq.~\eqref{eq:ir} is reached. Note that this
prescription for \IR{} extrapolation slightly differs and thus results
in marginal discrepancies in comparison with the one previously
employed by us in Ref.~\cite{Perez-Obiol:2020qjy}. In addition, \UV{}
corrections to Eq.~\eqref{eq:ir} can be significant and depend on
details of the nuclear and \YN{} interactions~\cite{Konig:2014hma}.
For interactions with non-local momentum regulators, like those used
in this work, \(\luv\) should significantly exceed \(\LamNN\) and
\(\LamYN\). A large-enough \luv{} scale can be identified by
performing calculations at a fixed \luv{}---by choosing appropriate
(\nm, \w{}) model-space parameters~\cite{Forssen:2017wei}---and
monitoring the dependence of results, such as the extrapolated energy
\(E_\infty(\luv)\) in Eq.~\eqref{eq:ir}, on the selected \luv{} scale.
We find that \(\luv = \qty{1200}{\MeV}\) is sufficient to achieve
\UV{} convergence; see also Ref.~\cite{Gazda:2022fte}. This is
demonstrated in Fig.~\ref{fig:l3h} where the \hyp{} g.s.\ energy
\(E(\hyp)^{\rm UV}\) is shown as a function of the IR length \lir{}
for several fixed values of the \UV{} scale
\(800 \leq \luv \leq \qty{1400}{MeV}\) (empty and filled symbols). The
NCSM calculations were performed with model space truncation up to
\(\nm = \nmmax\) using the \nnlosimstd{} and \loynstd{} interactions.
The energy is clearly converging with increasing the size of the model
space. The extrapolated values, \(E_{\infty}^{\rm UV}\), exhibit a
marginal dependence on the {\UV} scale of the \HO{} basis for
\(\luv \gtrsim \qty{1000}{MeV}\); see Table~\ref{tab:EGL3H} in
Sec.~\ref{sec:tbdr}. In fact, for \(E({\hyp})\) no extrapolation is
necessary since the \NCSM{} calculations performed at
\(\nm \approx \nmmax\) are converged with a precision of
few~\unit{\keV}. For example, by fixing
\(\luv= \qtylist[list-units=single, list-pair-separator={,}]{1000;
  1200}{\MeV}\), which implies
\(\w = \qtylist[list-units=single,list-pair-separator={,}]{7.299;
  10.510}{\MeV}\) for \(\nm=\nmmax\), we obtain
\(E^{1000, 1200}({\hyp}) =
\qtylist[list-units=single,list-pair-separator={,}]{-2.3807;-2.3814}{\MeV}\),
while the extrapolated infinite-space result estimated using
Eq.~\eqref{eq:ir} is
\(E_\infty^{1000, 1200} =
\qtylist[list-units=single,list-pair-separator={,}]{-2.385;
  -2.391}{\MeV}\). However, an adapted version of the IR-extrapolation
scheme~\eqref{eq:ir} will be applied in Sec.~\ref{sec:tbdr} to the
{\hyp} two-body {\pim} decay rate in Eq.~\eqref{eq:decayrate1} which
is found to have slightly different \NCSM{} model-space convergence
properties.

Most of the NCSM calculations in this work are performed for a
particular (hyper)nuclear Hamiltonian with fixed values of
\(\LamNN=\qty{500}{\MeV}\), \(\Tmax=\qty{290}{\MeV}\), and
\(\LamYN=\qty{600}{\MeV}\) cutoffs. A detailed analysis of theoretical
uncertainties associated with these cutoffs for relevant observables
is presented in Sec.~\ref{sec:unc}.

\section{Results}
\label{sec:results}
\subsection{Two-body \texorpdfstring{\(\pi^-\)}{pi minus} decay rate
  \texorpdfstring{\Gtbd{}}{Gamma(L3H to pi minus + 3He)}}
\label{sec:tbdr}

\noindent Employing the \NCSM{} \he{} and \hyp{} wave functions
calculated using the \nnlosimstd{} and \loynstd{} interactions, we now
proceed to calculate the two-body \pim{} decay rate {\Gtbd} in
Eq.~\eqref{eq:decayrate1}. While for \he{} the NCSM model-space
parameters were fixed at \(\nm=36\) and \(\w=\qty{14}{\MeV}\),
corresponding to a well-converged wave function, for \hyp{} we employ
wave functions generated up to \(\nm = \nmmax\) and \HO{} frequencies
\w{} which correspond to several fixed \UV{} cutoffs \luv{}, in order
to examine the \IR{} and \UV{} convergence of \Gtbd{}.
\begin{figure}[t]
  \centering
  \includegraphics[width=\columnwidth]{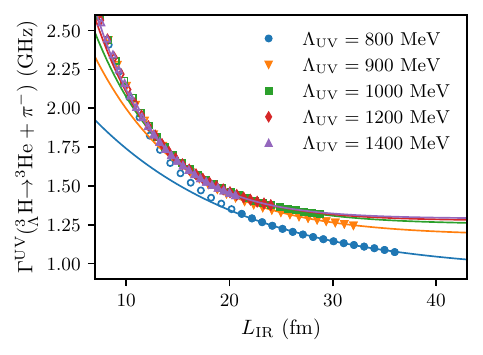}
  \caption{NCSM \hyp{} two-body \(\pi^-\) rates \Gtbd{} (empty and
    filled symbols) as functions of the \IR{} length \lir{} (up to
    \(\nm=\nmmax\)) calculated using the \nnlosimstd{} and \loynstd{}
    interactions for several fixed values of the \UV{} cutoff \luv{},
    together with their extrapolations (solid lines). The rates are
    evaluated using pion DW and considering the \SNN{}
    contributions. Only the points marked by filled symbols,
    corresponding to particle-stable \hyp{} configurations, are
    included in the fits. The NCSM model-space parameters for \he{}
    are fixed at \(\nm=36\) and \(\w=\qty{14}{\MeV}\), corresponding
    to a well-converged wave function.}
  \label{fig:decayrates}
\end{figure}
The calculated two-body rates \Gtbd{} are shown in
Fig.~\ref{fig:decayrates} as functions of the \IR{} \HO{} basis length
scale \lir{} for \luv=\qtylist[parse-numbers=false,
list-units=single]{800;900;1000;1200\text{,};1400}{\MeV}. The rates
include the effect of the distortion of the outgoing pion wave, as
well as contributions of the \(\Sigma\to N\,\pi\) transitions due to
\SNN{} admixtures in \hyp{}. They exhibit an exponential decrease with
\lir{} similar to, although slower than, the \hyp{} \gs{} energies in
Fig.~\ref{fig:l3h}. The solid lines in Fig.~\ref{fig:decayrates}
correspond to the \IR{} extrapolations into infinite NCSM model space.
The extrapolations were performed using an adapted version of
Eq.~\eqref{eq:ir} and included only the points marked by filled
symbols, corresponding to particle-stable \hyp{} configurations. The
extrapolated (\(\lir\to\infty\)) two-body \pim{} rates
\(\Gamma_\infty\), as well as the \hyp{} \gs{} energies \(E_\infty\),
are listed in Table~\ref{tab:EGL3H}. Both the extrapolated energies
and rates exhibit a rather small dependence on the \UV{} cutoff for
\(\num{1000} \lesssim \luv \lesssim \qty{1200}{\MeV}\). Also listed in
Table~\ref{tab:EGL3H} are the (extrapolated) \(\Lambda\) separation
energies in \hyp{} calculated using \(E_{\infty}\) and \IR{} converged
\nuc{2}{H} \gs{} energies \(E(\nuc{2}{H})\) at the corresponding
cutoff values \luv{}. As demonstrated in Fig.~\ref{fig:decayrates} and
Table~\ref{tab:EGL3H}, working at a fixed value of
\(\luv=\qty{1200}{\MeV}\) provides a good compromise between
minimization of the \UV{} corrections and reliable \IR{}
extrapolation. The small increase in \(E_{\infty}\) for
\(\luv = \qty{1400}{\MeV}\) indicates that higher-order \IR{}
corrections to Eq.~\eqref{eq:ir} become relevant for
\(\luv \gtrsim \qty{1400}{\MeV}\).
\begin{table}[h]
  \centering
  \caption{Extrapolated \hyp{} \gs{} energies \(E_\infty\) and
    two-body \(\pi^-\) decay rates \(\Gamma_\infty\) for several
    values of the \HO{} basis \UV{} scale, calculated using the
    \nnlosimstd{} and \loynstd{} interactions. The rates are evaluated
    using pion DW and considering the \SNN{} contributions. The
    values of the (extrapolated) \(\Lambda\) separation energies,
    \(\BL \equiv E({\nuc{2}{H}}) - E_\infty\), are obtained using
    converged deuteron \gs{} energies \(E(\nuc{2}{H})\) at the
    corresponding \luv{} scale.}
  \begin{ruledtabular}
    \begin{tabular}{l|rrrrr}
      \multicolumn{1}{l|}{\luv{} (\unit{\MeV})}
      & \multicolumn{1}{c}{\num{800}}
      & \multicolumn{1}{c}{\num{900}}
      & \multicolumn{1}{c}{\num{1000}}
      & \multicolumn{1}{c}{\num{1200}}
      & \multicolumn{1}{c}{\num{1400}} \\
      \(E_\infty\)  (\unit{\MeV})
      & \num{-2.287}
      & \num{-2.359}
      & \num{-2.385}
      & \num{-2.391}
      & \num{-2.388} \\
      \(E(\nuc{2}{H})\) (\unit{\MeV})
      & \num{-2.221}
      & \num{-2.224}
      & \num{-2.224}
      & \num{-2.224}
      & \num{-2.224} \\
      \BL{} (\unit{\MeV})
      & \num{0.066}
      & \num{0.135}
      & \num{0.161}
      & \num{0.167}
      & \num{0.164} \\
      \(\Gamma_\infty\) (\unit{\GHz})
      & \num{0.944}
      & \num{1.180}
      & \num{1.253}
      & \num{1.276}
      & \num{1.289}
    \end{tabular}
  \end{ruledtabular}
  \label{tab:EGL3H}
\end{table}

\subsection{Main contributions to \texorpdfstring{\Gtbd{}}{Gamma(L3H
    to pi minus + 3He)}}
\label{sec:decayrateapprox}

\noindent Out of all three-body channels \(\alpha\) in
Eq.~\eqref{eq:alpha}, only few contribute significantly to the wave
functions of \he{} and \hyp{}. The dominant component of the
hypertriton wave function is a deuteron-like core with
\(l_{12}= t_{12} =0, s_{12} = j_{12} = 1\) coupled to a \(\Lambda\)
hyperon (\(t_3 = 0\)) \(s\)-wave state. It contributes by
\(\approx \qty{96}{\percent}\) to the square of the norm of the \hyp{}
wave function. The analogous channel in \he{} accounts for
\(\approx \qty{47}{\percent}\), with additional
\(\approx\qty{47}{\percent}\) originating from the
\(l_{12}=s_{12}=l_3=0\) component with isospin \(t_{12}=1\).
Consequently, the largest contribution to the decay rate \Gtbd{} is
generated by the
\({}^{(2t_{12}+1)(2s_{12}+1)}{l_{12}}_{j_{12}} ={}^{13}S_1, l_3 = 0\)
components in the wave functions. By constraining the decay rate to
only these two components, \(\psi_{{}^{13}S_1,l_3=t_3=0}(\hyp)\) and
\(\psi_{{}^{13}S_1, l_3=0}'(\he)\), and considering {\pim} \PW{},
Eq.~\eqref{eq:decayrate2} reduces to a simpler expression,
\begin{equation}
  \label{eq:decayapprox}
  \begin{split}
    \Gamma^{\he}_{{}^{13}S_1}
    &= \frac{3}{4\pi}\frac{(G_Fm_\pi^2)^2\,M_{\he}\,q_{\pi}}{M_{\he}+E_{\pi}}
      \left(\mathcal{A}_\Lambda^2+\frac{1}{36}\frac{\mathcal{B}_\Lambda^2\,q_\pi^2}{\overline{M}^2_{\Lambda
      N}}\right) \\
    &\times \Big\lvert \int \mathrm{d}p_{12}'\,p_{12}'^2\int \mathrm{d}p_3'\,p_3'^2\int \mathrm{d}\cos \theta_{p_3'}\\
    &\times \psi_{^{13}S_1,l_3=t_3=0}(p_{12}',\lvert
      p_3'\hat{p}_3'+\tfrac{2}{3}\vec{q}_\pi\rvert)\\
    &\times \psi_{{}^{13}S_1,l_3=0}'(p_{12}',p_3') \Big\rvert^2.
  \end{split}
\end{equation}
For \(\luv = \qty{1200}{\MeV}\), Eq.~\eqref{eq:decayapprox} yields
\(\Gamma^{\he}_{{}^{13}S_1} = \qty{1.152}{\GHz}\) after extrapolation,
as listed in Table~\ref{tab:wfs}.
This \PW{} value differs merely by \(\approx\qty{4}{\percent}\) from
\qty{1.108}{\GHz} in the full calculation, partly due to cancellations
discussed below.
\begin{table}[htb]
  \centering
  \caption{Extrapolated partial decay rates
    \(\Gamma^{\rm UV}_{\infty}(\tbd{})\) (in~\unit{\GHz}) at
    \(\luv=\qty{1200}{\MeV}\) calculated using only the dominant \he{}
    and \hyp{} wave function components (first row) and considering
    only the \(\Lambda NN\) channels in \hyp{} (second row). Total
    rates including also the \SNN{} components in \hyp{} are in
    the last row. Probabilities are listed in percents and the rates
    in the third and fourth columns correspond to \pim{} {\PW} and
    {\DW}, respectively.}
  \begin{ruledtabular}
    \begin{tabular}{lcccc}
      Channel restrictions
      & \mbox{\(P(\nuc{3}{He})\)}
      & \mbox{\(P(\hnuc{3}{H})\)}
      & \mbox{\(\Gamma^{\mathrm{UV}}_{\mathrm{PW}}\)}
      & \mbox{\(\Gamma^{\mathrm{UV}}_{\mathrm{DW}}\)} \\
      \hline
      as in Eq.~\eqref{eq:decayapprox}
      & \num{46.81}
      & \num{95.83}
      & \num{1.152}
      & \\
      \(t_3(\hyp)=0\)
      & \num{100.00}
      & \num{99.59}
      & \num{1.243}
      & \num{1.426} \\
      none
      & \num{100.00}
      & \num{100.00}
      & \num{1.108}
      & \num{1.276}
    \end{tabular}
  \end{ruledtabular}
  \label{tab:wfs}
\end{table}

Subleading contributions to \Gtbd{} for both \pim{} {\PW} and {\DW}
are summarized in Table~\ref{tab:wfs}. The remaining channel
transitions due to \(\Lambda\to p\,\pi^-\) increase the rate only by
\(\approx\qty{8}{\percent}\), to \(\Gamma^{\he} = \qty{1.243}{\GHz}\)
in \PW{}. This increase originates mainly from the
\({}^{13}D_1, l_3 = 0\) components of the wave functions. In contrast,
including also the \SNN{} components in the \hyp{} wave function and
allowing for the \(\Sigma^-\to n\,\pim\) and \(\Sigma^0\to p\,\pim\)
transitions reduces the decay rate by \(\approx \qty{11}{\percent}\)
from \num{1.243} to \qty{1.108}{\GHz}, thus over-canceling the
\(\approx\qty{8}{\percent}\) increase from \num{1.152} to
\qty{1.243}{\GHz} owing to the non-\({}^{13}S_1\) \(\Lambda\)
amplitudes. Furthermore, replacing the outgoing \(\pim{}-\he{}\) {\PW}
by realistic \(\pim{}-\he{}\) {\DW} increases each of the listed \PW{}
rates by \(\approx \qty{15}{\percent}\). However, despite providing
substantial individual contributions, the combined effect of
accounting for the two new mechanisms considered in this work, namely
(i) including \SNN{} components of \hyp{} and (ii) introducing
\(\pim{}-\he{}\) \DW{}, is that they largely cancel each other with
barely \qty{3}{\percent} net increase of \Gtbd{} from \num{1.243} to
\qty{1.276}{\GHz}.

\begin{figure}[t]
  \centering
  \includegraphics[width=\columnwidth]{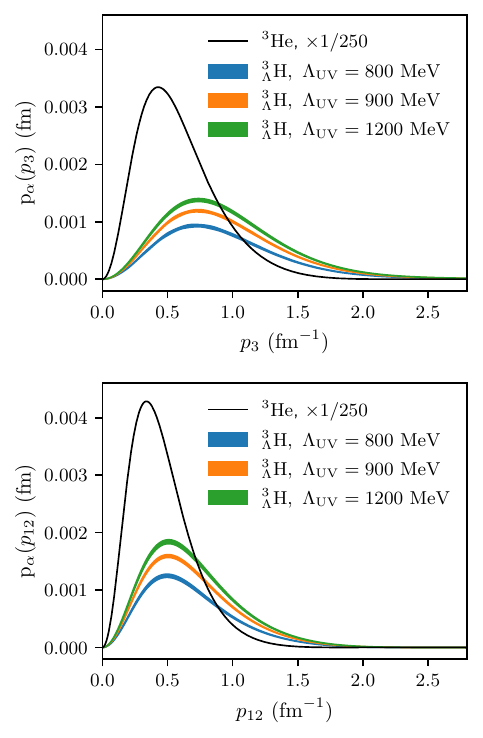}
  \caption{Probability densities
    \(\mathrm{p}_\alpha(p_3) \equiv \int \mathrm{d}p_{12}\, p_{12}^2
    \, p_3^2\, \lvert \psi_{\alpha}(p_{12},p_3)\rvert^2\) of the
    active baryon (\(N\) or \(\Sigma\)) relative momentum \(p_3\) for
    the \(\alpha = {}^{31}S_0, l_3{=}0\) \he{} and the dominant \SNN{}
    \(\alpha = {}^{31}S_0, l_3{=}0, t_3{=}1\) \hyp{} wave-function
    components (top panel), together with the corresponding
    distributions
    \(\mathrm{p}_\alpha(p_{12}) \equiv \int \mathrm{d}p_{3}\, p_{3}^2
    \, p_{12}^2\, \lvert \psi_{\alpha}(p_{12},p_3)\rvert^2\) of the
    {\NN} relative momentum \(p_{12}\) (bottom panel). The densities
    are calculated using the \nnlosimstd{} and \loynstd{}
    interactions. For \hyp{}, each band corresponds to a particular
    value of the \luv{} cutoff, while its width shows the variation
    with the model-space truncation \nm{} for
    \(60 \leq \nm \leq \nmmax\). For \he{}, the NCSM model-space
    parameters are fixed at \(\nm=36\) and \(\w=\qty{14}{\MeV}\),
    corresponding to a well-converged wave function. The
    \(\mathrm{p}_\alpha(p_{12})\) and \(\mathrm{p}_\alpha(p_3)\)
    distributions in \he{} are scaled by a factor of \(1/250\).}
\label{fig:sp12p3}
\end{figure}

The \(\approx\qty{11}{\percent}\) contribution to \Gtbd{} induced by a
tiny \(\approx\qty{0.4}{\percent}\) probability \SNN{} component of
\hyp{} is a bit unexpected, particularly when compared to the
\(\approx\qty{8}{\percent}\) comparable contribution of the
considerably stronger \({}^{13}D_1\) tensor component of order
\qty{3}{\percent} probability in the initial \hyp{} hypernucleus.
However, recall that whereas the \({^{13}D_1}\) \NN{} component in the
final \he{} that supports the \(\Lambda\to N\pi\) transition is of the
same order as in \hyp{}, the \({^{31}S_0}\) \NN{} component in \he{}
that supports the \(\Sigma\to N\pi\) transition is almost of
\qty{50}{\percent} probability. More importantly, since the two \PV{}
\(\Lambda\) and \(\Sigma\) weak decay amplitudes of
Eq.~\eqref{eq:weakop} interfere upon forming their summed absolute
value squared, even as small a \SNN{} admixture probability as
\qty{0.4}{\percent} in \hyp{} may affect considerably the calculated
\Gtbd{}, which was found here to be reduced by
\(\approx\qty{11}{\percent}\) from that evaluated disregarding the
\SNN{} admixture. This effect is further assisted by the overlap of
the relevant \SNN{} \hyp{} and \he{} wave-function components. Shown
in Fig.~\ref{fig:sp12p3} are the probability densities
\(\mathrm{p}_\alpha(p_{3}) \equiv \int \mathrm{d}p_{12}\, p_{12}^2 \,
p_3^2\, \lvert \psi_{\alpha}(p_{12},p_3)\rvert^2\) of the active
baryon (\(N\) or \(\Sigma\)) relative momentum \(p_3\) for the
\(\alpha = {}^{31}S_0, l_3{=}0\) \he{} and the dominant \SNN{}
\(\alpha = {}^{31}S_0, l_3{=}0, t_3{=}1\) \hyp{} wave-function
components, together with the associated \NN{} relative momentum
\(p_{12}\) distributions
\(\mathrm{p}_\alpha(p_{12}) \equiv \int \mathrm{d}p_3\, p_3^2 \,
p_{12}^2\, \lvert \psi_{\alpha}(p_{12},p_3)\rvert^2\). The \(\Sigma\)
momentum distribution in \hyp{} is localized around a higher value of
\(p_3\), closer to the peak of the active-nucleon momentum
distribution in \he{} compared to the \({}^{13}S_1,l_3=0\) \(\Lambda\)
and \(N\) momentum distributions shown in Fig.~\ref{fig:hypp3}.

\subsection{Dependence of \texorpdfstring{\Gtbd{}}{Gamma(L3H to
    pi minus + 3He)} on \texorpdfstring{\BL{}}{BL}}
\label{sec:luvtuning}

\begin{figure}[t]
  \centering
  \includegraphics[width=\columnwidth]{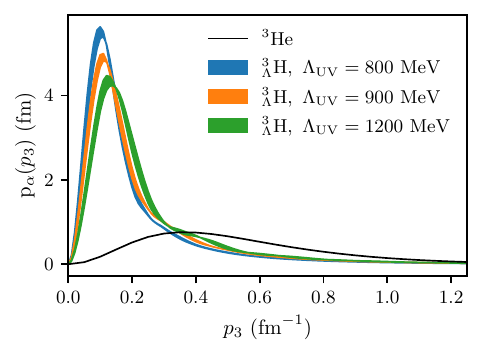}
  \caption{Probability densities \(\mathrm{p}_\alpha(p_3)\) as in
    Fig.~\ref{fig:sp12p3} of the dominant
    \(\alpha = {}^{13}S_1,l_3=0\) \he{} and
    \(\alpha = {}^{13}S_1,l_3=t_3=0\) \hyp{} wave-function components
    as functions of the active baryon (\(N\) or \(\Lambda\)) relative
    momentum \(p_3\). Note that
    \(\int \mathrm{d}p_3\, \mathrm{p}_\alpha(p_3) \approx
    \qty{47}{\percent}\) and \qty{96}{\percent} for \he{} and \hyp{},
    respectively.}
\label{fig:hypp3}
\end{figure}
\noindent While the \UV{} convergence of extrapolated separation
energies \(E_\infty(\hyp)\) and two-body decay rates
\(\Gamma_\infty({\tbd})\) for \(\luv \lesssim \qty{1200}{\MeV}\) was
not fully achieved (see Table~\ref{tab:EGL3H}), it is evident from
their \lir{} dependence, shown in Figs.~\ref{fig:2h3he} and
\ref{fig:l3h}, that the calculations still give meaningful,
sufficiently \IR{}-converged results. The missing \UV{} corrections to
Eq.~\eqref{eq:ir} depend only on short-range details of the employed
interactions which are truncated by the \luv{} cutoff. Nevertheless,
the correlation between \(E_\infty\) and \(\Gamma_\infty\) appear to
be robust, and their dependence on \luv{} can thus be exploited to
study the relationship of \BL{} and its associated \Gtbd{}. In
particular values of \(800 \lesssim \luv{} \lesssim \qty{1000}{\MeV}\)
provide meaningfully converged rates calculated using well-converged
\hyp{} wave functions at lower values of \BL{}. Indeed, it will be
shown in the following section that the relationship between the
separation energy and the two-body rate at a particular value of
\luv{} is consistent with the one obtained by varying the parameters
of the underlying nuclear and hypernuclear interactions.

The robust correlation between \BL{} and \Gtbd{} can be understood by
means of the approximate expression for the two-body decay rate in
Eq.~\eqref{eq:decayapprox}. In particular, it can be traced back to
the overlap between the
\(\psi_{^{31}S_0,l_3=t_3=0}(p_{12}',\lvert
p_3'\hat{p}_3'+\tfrac{2}{3}\vec{q}_\pi\rvert)\) \hyp{} and
\(\psi_{^{31}S_0,l_3=0}'(p_{12}',p_3')\) \he{} wave-function
components. The {\hyp} squared amplitude (probability distribution),
\(\mathrm{p}_\alpha(p_{12})\) (see Sec.~\ref{sec:decayrateapprox}),
with \(\alpha = {}^{13}S_1,l_3=t_3=0\), is almost independent of
\luv{} and peaked at \(p_{12}\approx \qty{0.25}{\per\femto\meter}\),
very close to the peak in the corresponding amplitude of the
\({}^{13}S_1\) component of the deuteron wave function. On the other
hand, the analogous \hyp{} squared amplitude
\(\mathrm{p}_\alpha(p_3)\) as a function of the relative momentum of
the loosely bound \(\Lambda\) peaks at
\(p_3\approx \qty{0.1}{\per\femto\meter}\), a value much lower than
the corresponding momentum \(p_3'\) of the active nucleon in \he{}. As
larger \luv{} cutoffs are considered, \(\mathrm{p}_\alpha(p_3)\) peaks
at larger momenta \(p_3\), the \(\Lambda\) hyperon becomes more bound,
and the overlap integral in Eq.~\eqref{eq:decayapprox} becomes larger.
The density distributions \(\mathrm{p}_\alpha(p_3)\) of the dominant
components \(\alpha\) of the \he{} and \hyp{} wave functions are shown
in Fig.~\ref{fig:hypp3} for
\(\luv =\qtylist[parse-numbers=false,
list-units=single]{800;1000\text{,};1200}{\MeV}\).

\subsection{Theoretical (hyper)nuclear-structure uncertainties in \texorpdfstring{\Gtbd{}}{Gamma(L3H to
    pi minus + 3He)}}
\label{sec:unc}

\noindent The nuclear and hypernuclear wave functions entering the
two-body \pim{} decay rate \Gtbd{} are associated with systematic
uncertainties resulting from the selection of calibration data, the
truncation of the chiral expansion, and possible regulator artifacts of
the employed \NN{}+\NNN{} and \YN{} interaction models.

\begin{figure}[tb]
  \centering
  \includegraphics[width=\columnwidth]{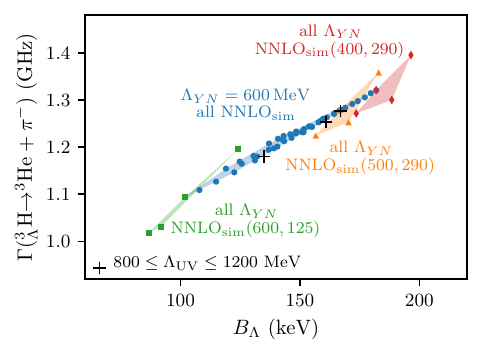}
  \caption{Variation of the extrapolated \(\Lambda\) separation
    energies in \hyp{} and decay rates \Gtbd{} with the \LamNN{},
    \Tmax{} and \LamYN{} cutoffs applied in the nuclear \nnlosim{} and
    LO \YN{} hypernuclear interactions. The rates are calculated
    including contributions from {\pim} DW and \SNN{}
    contributions for \(\luv=\qty{1200}{\MeV}\). The energies and
    rates are computed for a fixed value of \(\LamYN=\qty{600}{\MeV}\)
    and all 42 \nnlosim{} Hamiltonians (circles); and all cutoffs
    \(\LamYN = 550, 600, 650, \qty{700}{\MeV}\) for fixed values of
    \((\LamNN,\Tmax)=(600,125)~\unit{\MeV}\) (squares),
    \((500,290)~\unit{\MeV}\) (triangles), and
    \((400,290)~\unit{\MeV}\) (diamonds). Results obtained for
    different values of the \HO{} basis \UV{} scale
    \(800\leq\luv\leq\qty{1200}{\MeV}\) with
    \((\LamNN,\Tmax,\LamYN)=(500,290,600)~\unit{\MeV}\) are marked by
    black crosses. See text for details.}
  \label{fig:decayrateunc}
\end{figure}

In order to estimate the limits of theoretical precision of relevant
hypernuclear observables resulting from the nuclear model uncertainty,
we employ the whole \nnlosim{} family of 42 nuclear potentials; see
Sec.~\ref{sec:int}. In addition, we also quantify variation of the
observables related to the momentum regulator cutoff dependence of the
\loyn{} potential. In particular, we focus on the spread of the
predicted \(\Lambda\) separation energies \BL{} in \hyp{} and the
two-body {\hyp} decay rates \Gtbd{}. Results of this analysis are
summarized in Fig.~\ref{fig:decayrateunc} where the extrapolated
values of \BL{} and \Gtbd{} are obtained from NCSM calculations at
\(\luv=\qty{1200}{\MeV}\) for \nm{} up to \nmmax{}. The rates are
computed including the contributions from \pim{} DW and
\(\Sigma\to N\,\pi\) modes.
Results for a fixed value of the \YN{} interaction regulator cutoff
momentum \(\LamYN = \qty{600}{\MeV}\) and all 42 {\nnlosim} {\NN+\NNN}
potentials are presented by blue filled circles. The predicted
\(\Lambda\) separation energy varies strongly,
\(\num{100} \lesssim \BL \lesssim \qty{180}{\keV}\), and decreases
with increasing the regulator cutoff \LamNN{} and increases with
increasing \Tmax{}~\cite{Htun:2021jnu,Gazda:2022fte}. The two-body
\(\pi^-\) rate is strongly correlated with \BL{} and varies in the
range \(1.1 \lesssim \Gtbd \lesssim \qty{1.3}{\GHz}\). Surprisingly,
this correlation is perfectly in line with the results obtained for
different values of the \HO{} basis \UV{} scale listed in
Table~\ref{tab:EGL3H} and included in Fig.~\ref{fig:decayrateunc} for
\(800\leq\luv\leq\qty{1200}{\MeV}\) (black crosses). Also shown in
Fig.~\ref{fig:decayrateunc} is the variation of \BL{} and \Gtbd{} with
the {\YN} regulator cutoff momentum \LamYN{}. We selected only 3
\nnlosim{} Hamiltonians, two of which give the lowest and highest
separation energies and rates, together with {\nnlosimstd}. The
energies and rates are calculated for
\(\LamYN =
\qtylist[parse-numbers=false,list-units=single]{550;600;650\text{,};700}{\MeV}\)
and fixed values of
\((\LamNN,\Tmax) = \qty[parse-numbers=false]{600,125}{\MeV}\)
(squares), \(\qty[parse-numbers=false]{(500,290)}{\MeV}\) (triangles),
and \(\qty[parse-numbers=false]{(400,290)}{\MeV}\) (diamonds). In each
group, the largest separation energy always occurs for
\(\LamYN = \qty{550}{\MeV}\), decreases monotonically with \LamYN{}
for \(550\leq\LamYN\leq\qty{650}{\MeV}\), and increases for
\(650\leq\LamYN\leq\qty{700}{\MeV}\). Note that the resulting combined
spread of binding energies for all considered
\((\LamNN,\Tmax,\LamYN)\) cutoff combinations
\(\Delta \BL \approx \qty{100}{\keV}\) is essentially of the same
order as the experimental uncertainty, while
\(\Delta \Gtbd \approx \qty{0.4}{\GHz}\).

Very recently, a comparable, although smaller, variation of \(\BL\) in
\hyp{} with the \LamNN{} regulator cutoff momentum using higher-order
\chieft{} \NN{} interactions and \loynstd{} was reported in
\cite{Le:2023bfj}. In addition, the sensitivity of \BL{} to the
employed nuclear interactions was also found there to decrease
considerably when higher-order \YN{} potentials were employed.

\subsection{Hypertriton lifetime}
\label{sec:res:lifetime}

\noindent To evaluate the total \hyp{} decay rate and lifetime we
employ the strategy described in detail in Sec.~\ref{sec:lifetime}.
Using the {\IR}-extrapolated two-body \pim{} decay rate
\(\Gamma_\infty^{1200}=\qty{1.276}{\GHz}\), associated with
\(\BL^{1200}=\qty{167}{\keV}\), we evaluate the inclusive \pim{}
\hyp{} decay rate \(\Gamma_{\pim}(\hyp)\) by means of the measured
world-average \BC{} branching ratio
\(R_3 = \Gtbd{} / \Gamma_{\pi^-}(\hyp) =
\BRval\)~\cite{Keyes:1973two}. By multiplying \(\Gamma_{\pi^-}(\hyp)\)
by the \(\DeltaI = \frac12\) factor of \(\frac32\) to include the
neutral pion \hyp{} decay channels,
\(\Gamma_\pi(\hyp) = \frac32\, \Gamma_\pim(\hyp)\), we derive the
hypertriton lifetime corresponding to all pionic decay modes as
\(\tau_\pi(\hyp) = 1/\Gamma_\pi(\hyp) =
\qty[parse-numbers=false]{183(21)}{\ps}\). Here and in what follows,
the quoted lifetime
uncertainty is purely statistical, arising from that of \(R_3\). The
nonmesonic \(\Lambda N\to NN\) and pion true-absorption
\(\pi + NN\to NN\) contributions to the rate of \qty{1.5}{\percent}
and \qty{0.8}{\percent}, respectively, further shorten the \hyp{}
lifetime. Accounting for their \qty{2.3}{\percent} combined increase
of the rate, we obtain
\(\thyp{} =\qty[parse-numbers=false]{179(20)}{\ps}\). This value is
considerably shorter, by
\(\approx \qty[parse-numbers=false]{32(8)}{\percent}\), than the
\(\Lambda\) lifetime in free space,
\(\tl = \qty[parse-numbers=false]{263(2)}{\ps}\). However,
significantly longer \hyp{} lifetime values, listed in
Table~\ref{tab:thyp}, are obtained by repeating the same procedure for
the two-body \pim{} decay rates from Table~\ref{tab:EGL3H} calculated
at lower values of \HO{} basis \UV{} scale,
\(\luv{} = \qtylist[parse-numbers=false, list-units=single]{800;900\text{,};1000}{\MeV}\). As
argued in Secs.~\ref{sec:luvtuning} and~\ref{sec:unc}, these rates
are associated with smaller values of \(\Lambda\) separation energies
\BL{}. It should be pointed out that for each of the derived \thyp{}
listed in Table~\ref{tab:thyp} we have tacitly assumed the same value
of branching ratio \(R_3\), taken from
experiment~\cite{Keyes:1973two}.
\begin{table}[t]
  \caption{Extrapolated \(\Lambda\) separation energies in \hyp{} and
    associated lifetimes \thyp{} (both in \unit{ps} and \tl{},
    with their \(R_3\)-induced uncertainties)
    calculated using the \nnlosimstd{} and \loynstd{} interactions for
    several values of the \HO{} basis \UV{} scale, and extrapolated to
    the STAR Collaboration~\cite{STAR:2019wjm} reported value of \BL.
    The lifetimes are evaluated using
    \(R_3 = \BRval{}\)~\cite{Keyes:1973two}, the \(\DeltaI = \frac12\)
    rule, and include a \qty{2.3}{\percent} correction from nonmesonic
    decay rate.}
  \label{tab:thyp}
  \begin{center}
    \begin{ruledtabular}
      \begin{tabular}{l|ccccc}
        \luv{} ({\unit{\MeV}}) & \num{800}      & \num{900}      & \num{1000}    & \num{1200}    & \\
        \BL{} (\unit{{\keV}})  & \num{66}       & \num{135}      & \num{161}     & \num{167}     & \num{410} \\
        \thyp{} (\unit{\ps})   & \num{242(28)}  & \num{193(22)}  & \num{182(21)} & \num{179(20)} & \num{158(18)} \\
        \thyp{} (\tl)          & \num{0.92(10)} & \num{0.73(8)}  & \num{0.69(8)} & \num{0.68(8)} & \num{0.60(7)}
      \end{tabular}
    \end{ruledtabular}
  \end{center}
\end{table}
Note that we obtain for the least-bound \hyp{} case at
\(\luv=\qty{800}{\MeV}\), a \thyp{} value which is shorter than \tl{}
only by less than \(\approx\qty{10}{\percent}\).

Remarkably, this value,
\(\tau^{800}(\hyp) = \qty[parse-numbers=false]{242(28)}{\ps}\), is
consistent with the most recent ALICE
Collaboration's~\cite{ALICE:2022sco} reported lifetime
\(\tau^{\text{ALICE}}(\hyp)=\qty[parse-numbers=false]{253(11)(6)}{\ps}\)
while its corresponding \(\Lambda\) separation energy
\(\BL^{800} = \qty{66}{\keV}\) falls comfortably within the ALICE
reported separation energy interval
\(\BL^{\text{ALICE}} = \qty[parse-numbers=false]{102(63)(67)}{\keV}\).
Within its \(R_3\)-induced uncertainty, \(\tau^{800}(\hyp)\) is also
consistent with lifetime value derived in a fully three-body
calculation in Ref.~\cite{Kamada:1997rv}. Similarly, the lifetime for
\(\luv = \qty{1000}{\MeV}\) listed in Table~\ref{tab:thyp} agrees well
within measurement uncertainties with the HypHI Collaboration's
lifetime value
\(\tau^{\mathrm{HypHI}}(\hyp) =
\qty[parse-numbers=false]{183^{+42}_{-32}(37)}{\ps}\). In order to
compare with STAR Collaboration's reported lifetime
\(\qty[parse-numbers=false]{142^{+24}_{-21}(29)}{\ps}\)~\cite{STAR:2017gxa}
and their own value of \(\Lambda\) separation energy
\(\BL =
\qty[parse-numbers=false]{0.41(12)(12)}{\MeV}\)~\cite{STAR:2019wjm},
we expand \Gtbd{} in powers of \(\sqrt{B_\Lambda}\) as
\(a \sqrt{\BL}+b\,\BL\) and fix the two expansion coefficients by
fitting to the \(\BL^{\mathrm{UV}}\) and
\(\Gamma_{\infty}^{\mathrm{UV}}\) from Table~\ref{tab:EGL3H} for
\(800 \le \luv \le \qty{1200}{\MeV}\). When extrapolated to
\(\BL = \qty{410}{\keV}\), we obtain \(\thyp = \qty{158(18)}{\ps}\).
Had we considered STAR Collaboration's own value of
\(R_3^{\text{STAR}} = 0.32(5)(8)\)~\cite{STAR:2017gxa}, the estimated
central value of
\(\thyp{} = \qty[parse-numbers=false]{145(51)}{\GHz}\) is almost
coincident with STAR Collaboration's \thyp{} central value. Moreover,
the most recent STAR Collaboration's reported lifetime
\qty[parse-numbers=false]{221(15)(19)}{\ps}~\cite{STAR:2021orx} is
consistent with lower \(\Lambda\) separation energies
\(\BL \approx \qty{90}{\keV}\) using \(R_3=\BRval\), which are also
plausible given the large measurement uncertainty in \BL{}.

Altogether, given the strong dependence of \thyp{} on \BL{} and
considering the large experimental uncertainty in \BL, none of the
recent {\RHI} reported \thyp{} values can be excluded, but rather can
be associated with its own underlying value of \BL{}.

\section{Conclusions}
\label{sec:conclusions}

\noindent We performed a new microscopic calculation of the
hypertriton \(\pi^-\) two-body decay rate \Gtbd{} employing \hyp{} and
\he{} three-body wave functions generated by the ab~initio \NCSM{}
approach using realistic chiral \YN{} and \NN{}+\NNN{} interactions as
the only input. Employing the \(\DeltaI=\frac12\) rule and the
experimental value of branching ratio
\(R_3=\Gtbd{}/\Gamma_{\pi^-}(\hyp)\) to include the remaining
\(\pi^0\) and three-{} plus four-body \hyp{} decay channels, we were able to
deduce the \hyp{} lifetime \thyp{}.

The following are the main findings and conclusions of this study:
\begin{enumerate}[(i)]
\item \textit{Pionic {\FSI}.} Considering the distortion of the
  emitted pion wave due to its strong-interaction attractive
  final-state interaction with the \nuc{3}{He} nucleus increases
  \Gtbd{} by \(\approx\qty{15}{\percent}\). A somewhat stronger
  enhancement of \(\approx\qty{18}{\percent}\) was found very recently
  in a pionless {\eft} calculation~\cite{Hildenbrand:2023rkc}.
\item \textit{Effect of \SNN{} admixtures.} Despite the negligible
  \SNN{}-component admixture (\(\lesssim \qty{0.5}{\percent}\)) in the
  \hyp{} wave function, the \(\Sigma\to N\,\pi\) transitions reduce
  \Gtbd{} by \(\approx\qty{11}{\percent}\) due to interference
  effects. It would be interesting to evaluate this effect using other
  \chieft{} \YN{} potential versions than the LO version used here,
  thus adding the measured value of \thyp{} as a useful constraint on
  such \YN{} potentials.
\item \textit{Relationship of {\BL} and \Gtbd{}.} The two-body \pim{}
  decay rate is found to be very sensitive to the \(\Lambda\) separation
  energy {\BL} in \hyp{}, the value of which is rather poorly known
  experimentally and also suffers from large theoretical uncertainties.
\item \textit{Hypertriton lifetime.} Using the \nnlosimstd{} and
  \loynstd{} nuclear and hypernuclear interactions which yield
  \(\BL^{\mathrm{th}}(\hyp) = \qty{167}{\keV}\)---consistent with the
  the world-average measured \(\BL^{\mathrm{exp.}}(\hyp) =
  \qty[parse-numbers=false]{164(43)}{\keV}\)~\cite{Hypweb}---to
  calculate \Gtbd{} and employing the branching ratio
  \(R_3=\BRval\)~\cite{Keyes:1973two}, together with the empirical
  \(\DeltaI = \frac12\) rule, we obtained the hypertriton lifetime
  \(\thyp = \qty{179(20)}{\ps} \approx 0.7(1)\,\tl{}\).
\item \textit{Magnitude of theoretical nuclear and hypernuclear
    structure uncertainties.} The combined spread of \(\Lambda\)
  separation energies resulting from variation of the \((\LamNN, \Tmax,
  \LamYN)\) \chieft{} interaction cutoffs is found to be \(90 \lesssim
  \BL \lesssim \qty{190}{\keV}\), while the spread in calculated
  two-body \(\pi^-\) rates \(1.0 \lesssim \Gtbd \lesssim
  \qty{1.4}{\GHz}\). This implies \(160(20) \lesssim \thyp \lesssim
  \qty[parse-numbers=false]{230(30)}{\ps}\) for \(R_3 = \BRval\).
\item \textit{Comparison with recent {\RHI} measurements.} The
  lifetime \thyp{} varies strongly with \BL{}. It is then not possible
  to exclude any of the distinct {\RHI} \thyp{} measured values, but
  rather relate the lifetime with its own underlying value of \BL{}. We note the
  good agreement between the very recent ALICE measured lifetime value
  \(\tau^{\text{ALICE}}(\hyp) =
  \qty[parse-numbers=false]{253(11)(6)}{\ps}\) associated with the ALICE
  measured {\BL} value \(B^{\text{ALICE}}_{\Lambda} =
  \qty[parse-numbers=false]{102(63)(67)}{\keV}\)~\cite{ALICE:2022sco}
  and the lifetime value
  \(\thyp=\qty[parse-numbers=false]{242(28)}{\ps}\) computed at the
  lowest value \(\BL = \qty{66}{\keV}\) reached by us. Nevertheless,
  only future experiments expected at MAMI, JLab, J-PARC, and CERN will
  hopefully pin down {\BL} with a better precision than \qty{50}{\keV}
  and lead to a resolution of the {\hyp} lifetime puzzle.
\end{enumerate}

\acknowledgments

We are grateful to Petr Navr\'{a}til for helpful advice on extending
the nuclear \NCSM{} codes to hypernuclei; to Johann Haidenbauer
and Andreas Nogga for providing us with the input \LO{}
Bonn--J\"{u}lich \(\YN\) potential; and to Andreas Ekstr\"{o}m for
providing us the \nnlosim{} \NN{}+\NNN{} potentials used in the
present work. We also thank Assumpta Parre\~{n}o and \`{A}ngels Ramos
for valuable discussions. This work was supported by the European
Union’s Horizon 2020 research and innovation program under Grant
Agreement No.~824093 (E.F., A.G., and D.G.); by the Czech Science
Foundation GA\v{C}R Grant No.~22-14497S (D.G); and by Secretaria
d’Universitats i Recerca del Departament d’Empresa i Coneixement de la
Generalitat de Catalunya, cofunded by the European Union Regional
Development Fund within the ERDF Operational Program of Catalunya
(project QuantumCat, ref.\ 001-P-001644) (A.P.-O.). Some of the
computational resources were supplied by IT4Innovations Czech National
Supercomputing Center supported by the Ministry of Education, Youth
and Sports of the Czech Republic through the e-INFRA~CZ (ID:90140).

\appendix
\section{Weak-decay operator matrix elements}
\label{sec:matrixelements}

\noindent In order to evaluate the matrix element of the transition
operator \eqref{eq:weakop} in Eq.~\eqref{eq:decayrate2} between the
nuclear, hypernuclear \eqref{eq:wfpw}, and pionic \eqref{eq:phipi}
wave functions we proceed by decoupling the spin and isospin states as
\begin{equation}
  \label{eq:matrixelementdw}
  \begin{split}
    &\Braket{\Psi_{\he}\,\phi_\pi|\hat{O}|\Psi_{\hyp}}\\
    &= \sum_{\alpha'}\int \mathrm{d}p_{12}'\,{p_{12}'}^2\,\mathrm{d}p_3'\,{p_3'}^2
    \psi_{\alpha^\prime}(p_{12}',p_3')\\
    &\times\sum_\alpha\sum_{m_{j_{12}},m_{l_3}',m_{l_3}}
    \delta_{l_{12}l_{12}'} \delta_{s_{12}s_{12}'}
    \delta_{j_{12}j_{12}'} \delta_{m_{j_{12}}m_{j_{12}}'}\\
    &\times \delta_{M^\prime M}
    \delta_{m_{s_3}m_{s_3}'}
    \cgC{j_{12}}{j_3'}{J'}{m_{j_{12}}}{M'{-}m_{j_{12}}}{M'}
    \cgC{j_{12}}{j_3}{J}{m_{j_{12}}}{M{-}m_{j_{12}}}{M}\\
    &\times \cgC{l_3'}{s_3'}{j_3'}{m_{l_3}'}{M'{-}m_{j_{12}}{-}m_{l_3}'}{M'{-}m_{j_{12}}}
    \cgC{l_3}{s_3}{j_3}{m_{l_3}}{M{-}m_{j_{12}}{-}m_{l_3}}{M{-}m_{j_{12}}}\\
    &\times
    \sqrt{2}\,G_Fm_\pi^2 \left[\left(
        \mathcal{A}_\Lambda+\frac{\mathcal{B}_\Lambda\,q_\pi}{2\overline{M}_{\Lambda
            N}}(-1)^{m_{s_3}'-\frac12}\right)\delta_{t_3 0}\right.\\
    &\left.+\frac{1}{2}\mathcal{A}_{\Sigma^-n\pi^-}\delta_{t_31} \right]
    \int \mathrm{d}p_3\,p_3^2\,\psi_{\alpha}(p_{12}',p_3)\,r^{-3} \\
    &\times \sum_{l_\pi}\int
    \mathrm{d}\hat{p}_3'\,\mathrm{d}\hat{p}_3
    Y_{l_3'm_{l_3}'}^*(\hat{p}_3')
    Y_{l_3 m_{l_3}}(\hat{p}_3)\\
    &\times Y_{l_\pi 0}^*(\widehat{\vec{p}_3-\vec{p}_3^{\,\prime}})
    \,\tilde{\phi}_{l_\pi 0}\!\left(\tfrac{3}{2}\lvert\vec{p}_3-\vec{p}_3^{\,\prime}\rvert\right),
  \end{split}
\end{equation}
where \(m_{s_3} = M - m_{j_{12}}- m_{l_3}\),
\(m_{s_3}' = M' - m_{j_{12}}' -m_{l_3}'\),
\(C_{j_1,j_2;j_3,m_1}^{m_2; m_3}\equiv \braket{j_1 m_1 j_2 m_2| j_3
  m_3}\) are the \CG{} coefficients, and the spin matrix element was
evaluated using
\begin{equation*}
  \Braket{m_{s_3}' | \vec{\sigma}\cdot q_\pi\hat{z} |  m_{s_3}}=
  (-1)^{m_{s_3}'-\frac12}\delta_{m_{s_3}m_{s_3}'}q_\pi.
\end{equation*}
The nonprimed and primed quantities are associated with the initial
and final states, respectively. Note that we assume the pion momentum
\(\vec{q}_\pi\) in the \(\hat{z}\) direction and the expansion of the
pion wave in spherical harmonics has contributions only from
\(m_\pi = 0\). Consequently, \(m_{l_3}=m_{l_3'}\) and the matrix element
only depends on the difference of azimuthal angles
\(\theta_{p_3}-\theta_{p_3'}\) of \(\hat{p}_3\) and \(\hat{p}_3'\). We can
integrate out one of them, for example by fixing \(\theta_{p_3}=0\) and
replacing
\begin{equation}
  \label{eq:repl}
  \int \mathrm{d}\hat{p}_3'\to 2\pi\int \mathrm{d}\cos{\theta_{p_3'}}.
\end{equation}

The matrix element~\eqref{eq:matrixelementdw} simplifies considerably
when the {\pim} wave function is approximated by a {\PW},
\(\tilde{\phi}_{\pi;\vec{q}_\pi}(\vec{k}) =
\braket{\vec{k}|q_\pi\hat{z}} = 1/k^2 \delta^{(3)}(\vec{k} - q_\pi
\hat{z})\). In this case, the matrix element evaluates to
\begin{equation}
  \label{eq:matrixelementpw}
  \begin{split}
    &\Braket{\Psi_{\he}\,\phi_\pi | \hat{O} | \Psi_{\hyp}}\\
    &= \sum_{\alpha'}\int
    \mathrm{d} p_{12}'\,{p_{12}'}^2\,\mathrm{d} p_3'\,{p_3'}^2\,
    \psi_{\alpha}'(p_{12}',p_3')
    \sum_{\alpha} \sum_{m_{j_{12}},m_{l_3}} \\
    &\times
    \delta_{l_{12}l_{12}'} \delta_{s_{12}s_{12}'}
    \delta_{j_{12}j_{12}'} \delta_{m_{j_{12}}m_{j_{12}}'}
    \delta_{M',M} \delta_{m_{s_3}m_{s_3}'} \\
    &\times \delta_{m_{l_3}m_{l_3}'}
    \cgC{j_{12}}{j_3'}{J'}{m_{j_{12}}}{M'-m_{j_{12}}}{M'}
      \cgC{j_{12}}{j_3}{J}{m_{j_{12}}}{M-m_{j_{12}}}{M} \\
    &\times
    \cgC{l_3'}{s_3'}{j_3'}{m_{l_3}'}{M'-m_{j_{12}}-m_{l_3}'}{M'-m_{j_{12}}}
    \cgC{l_3}{s_3}{j_3}{m_{l_3}}{M-m_{j_{12}}-m_{l_3}}{M-m_{j_{12}}}
    \\
    &\times
    \sqrt{2}\,G_Fm_\pi^2
    \left[\left(
      \mathcal{A}_\Lambda+\frac{\mathcal{B}_\Lambda\,q_\pi}{2\overline{M}_{\Lambda
          N}}(-1)^{m_{s_3}'-\frac12}\right)\delta_{t_30}\right. \\
    &+\left.\frac{1}{2}\mathcal{A}_{\Sigma^-n\pi^-}\delta_{t_31}
      \right]
      2\pi \int \mathrm{d} \cos{\theta_{p_3'}}\,
      Y_{l_3 'm_{l_3}'}^*(\hat{p}_3') \\
    &\times
    Y_{l_3m_{l_3}}(\widehat{\vec{p}_3\,'+\tfrac{2}{3}\,q_{\pi}\hat{z}})\,
    \psi_{\alpha}(p_{12}',\lvert\vec{p}_3\,'+\tfrac{2}{3}\,q_{\pi}\hat{z}\rvert),
  \end{split}
\end{equation}
where the nonprimed and primed quantities are again associated with
the initial and final states, respectively. Here, the azimuthal part
of \(\hat{p}_3'\) has been fixed to zero since there is no dependence
on it and the factor \(2\pi\) comes from the
replacement~\eqref{eq:repl}.

\section{Fourier transform of the distorted
  \texorpdfstring{\(\pi^-\)}{pi minus} wave function}
\label{app:ft}

\noindent The distorted pion wave function is typically obtained by
solving Schr\"odinger or Klein--Gordon equation with
\(\pi^-\)--nuclear optical potentials in position space. The
momentum-space representation, entering the matrix element in
Eq.~\eqref{eq:decayrate1}, is obtained by its Fourier transform.

For a function expanded in spherical harmonics \(Y_{l m}(\hat{r})\) as
\begin{equation}
  \label{eq:expf}
  f(\vec{r})= \sum_{l,m}R_{lm}(r) Y_{l m}(\hat{r}),
\end{equation}
where the coefficients \(R_{lm}(r)\) contain its radial dependence, we
define the Fourier transform by
\begin{equation}
  \label{eq:fexpf}
  \begin{split}
    \mathcal{F}\left[f(\vec{r})\right](\vec{k}) &\equiv \frac{1}{(2\pi)^3}\int
    \mathrm{d}^3r\,e^{-i\vec{k}\cdot\vec{r}}f(\vec{r}) \\
    &= \frac{1}{2\pi^2}\sum_{l,m}(-i)^l Y_{lm}(\hat{k})\\
    &\times \int_0^\infty \mathrm{d} r
    r^2 j_l(k r)R_{lm}(r).
  \end{split}
\end{equation}
Here, the exponential function was expanded in terms of spherical
harmonics and spherical Bessel functions \(j_l(k r)\) using
\begin{equation}
  e^{-i\vec{k}\cdot\vec{r}}=4\pi\sum_{l,m}(-i)^{l}j_l(k r)Y_{lm}^*(\hat{k})Y_{lm}(\hat{r}).
\end{equation}
Note that for a \(\pi^-\) {\PW},
\(f(\vec{r}) = e^{i\vec{q}_\pi\cdot\vec{r}}\), the coefficient
functions in Eq.~\eqref{eq:expf} become
\begin{equation}
  \label{eq:rpw}
  R_{lm}^{(\mathrm{PW})}(r) = 4\pi\,i^l Y_{lm}^*(\hat{q}_\pi)j_l(q_\pi r)
\end{equation}
and we recover the momentum-space \(\pi^-\) {\PW}
\begin{equation}
  \label{eq:frpw}
  \mathcal{F}\left[e^{i\vec{q}_\pi\cdot\vec{r}}\right](\vec{k}) =
  \frac{1}{k\, q_\pi}\delta(\hat{k}-\hat{q}_\pi)\,\delta(k-q_\pi),
\end{equation}
in line with our normalization,
\(\braket{\vec{k} | \vec{q}_\pi} = 1/k^2\,
\delta^{(3)}(\vec{k} - \vec{q}_\pi)\).
In case of \(\pi^-\) {\DW}, the spherical Bessel functions
in Eq.~\eqref{eq:rpw} are replaced by the partial-wave components
\(\phi_l(q_\pi; r)/(q_\pi\, r)\) of the coordinate-space \(\pi^-\) wave
function. The Fourier transform of the {\DW} can be expressed
as
\begin{multline}
  \label{eq:fdw}
  \mathcal{F}\Big[ \sum_{l,m} Y_{lm}(\hat{r}) R_{l m}^{(\mathrm{DW})}(r) \Big](\vec{k}) \\ = \sum_{l,m}\tilde{\phi}_{l m}(k)\,Y_{l m}(\hat{k}),
\end{multline}
where
\begin{equation}
  \label{eq:rdw}
  R_{lm}^{(\mathrm{DW})}(r) = 4\pi\,i^l\, Y_{lm}^*(\hat{q}_\pi)\frac{1}{q_\pi\, r}\,\phi_l(q_\pi; r),
\end{equation}
and the momentum-space \(\pi^-\) partial waves
\begin{equation}
  \label{eq:ftpion}
  \begin{split}
    \tilde{\phi}_{l m}(k) &= Y_{l
      m}^{*}(\hat{q}_\pi) \\ &\times \frac{2}{\pi\,q_\pi}
    \int_0^{R_{\mathrm{max}}} {\rm d}r\, r\, j_l(k r)\,\phi_l(q_\pi;r)
  \end{split}
\end{equation}
then enter the matrix elements of the weak-decay operator in
Eq.~\eqref{eq:matrixelementdw}. During evaluation of the matrix
elements, the pion momentum is assumed to point in the
\(z\)-direction, \(\vec{q}_\pi=q_\pi \hat{z}\), and the spherical
harmonic in Eq.~\eqref{eq:ftpion} reduces to \(Y_{l 0}^{*}(\hat{z})\).
Note that a finite value of \(R_{\mathrm{max}}\) in
Eq.~\eqref{eq:ftpion} introduces a certain limit on the momentum scale
that \(\tilde{\phi}_{l m}(k)\) can probe. The function
\(\tilde{\phi}_{l m}(k)\) oscillates with frequencies proportional to
\(R_{\mathrm{max}}\). For example, in the case when
\(\phi_l(q_\pi;r)\) is a {\PW} and \(\tilde{\phi}_{l m}(k)\) is an
approximation to \(\delta(k - q_\pi)\), the amplitude of these
oscillations should be small enough to suppress any structure of the
wave functions multiplying \(\delta(k - q_\pi)\) in the matrix
elements at this scale. In the numerical implementation, the value
\(R_{\mathrm{max}}=\qty{60}{fm}\) for the upper bound of the integral
in Eq.~\eqref{eq:ftpion} was found sufficiently large to capture the
low-momentum structure of the \(\pi^-\) wave function. On the other
hand, the momentum grid used in computing the matrix elements has been
chosen fine enough, \(\Delta k = \qty{0.0025}{fm^{-1}}\), such that
the oscillations in \(\tilde{\phi}_{l m}(k)\) are well resolved.
We verified that the calculated decay rates were stable with
increasing \(R_{\mathrm{max}}\) and decreasing \(\Delta k\).


\begin{thebibliography}{51}

\bibitem{Perez-Obiol:2020qjy}
  A.~P\'erez-Obiol, D.~Gazda, E.~Friedman, and A.~Gal,
  Revisiting the hypertriton lifetime puzzle,
  \href{https://doi.org/10.1016/j.physletb.2020.135916}{Phys. Lett. B \textbf{811}, 135916 (2020)},
  \href{https://arxiv.org/abs/2006.16718}{arXiv:2006.16718~[nucl-th]}

\bibitem{ALICE:2022sco}
  S.~Acharya et al.\ (ALICE Collaboration),
  Measurement of the lifetime and $\Lambda$ separation energy of ${}^{3}_{\Lambda}\mathrm H$,
  \href{https://doi.org/10.1103/PhysRevLett.131.102302}{Phys. Rev. Lett. \textbf{131}, 10, 102302 (2023)},
  \href{https://arxiv.org/abs/2209.07360}{arXiv:2209.07360~[nucl-ex]}

\bibitem{Gal:2016boi}
  A.~Gal, E.~V. Hungerford, and D.~J. Millener,
  Strangeness in nuclear physics,
  \href{https://doi.org/10.1103/RevModPhys.88.035004}{Rev. Mod. Phys. \textbf{88}, 035004 (2016)},
  \href{https://arxiv.org/abs/1605.00557}{arXiv:1605.00557~[nucl-th]}

\bibitem{Le:2023bfj}
  H.~Le, J.~Haidenbauer, U.~G.~Mei{\ss}ner and A.~Nogga,
  Separation energies of light $\Lambda$ hypernuclei and their theoretical uncertainties,
  \href{https://doi.org/10.1140/epja/s10050-023-01219-w}{Eur. Phys. J. A \textbf{60}, 3 (2024)},
  \href{https://arxiv.org/abs/2308.01756}{arXiv:2308.01756~[nucl-th]}

\bibitem{Logoteta:2019utx}
  D.~Logoteta, I.~Vidana, and I.~Bombaci,
  Impact of chiral hyperonic three-body forces on neutron stars,
  \href{https://doi.org/10.1140/epja/i2019-12909-9}{Eur. Phys. J. A \textbf{55}, 207 (2019)},
  \href{https://arxiv.org/abs/1906.11722}{arXiv:1906.11722~[nucl-th]}

\bibitem{Gerstung:2020ktv}
  D.~Gerstung, N.~Kaiser, and W.~Weise,
  Hyperon\textendash{}nucleon three-body forces and strangeness in neutron stars,
  \href{https://doi.org/10.1140/epja/s10050-020-00180-2}{Eur. Phys. J. A \textbf{56}, 175 (2020)},
  \href{https://arxiv.org/abs/2001.10563}{arXiv:2001.10563~[nucl-th]}

\bibitem{Friedman:2022bpw2023ucs}
E.~Friedman and A.~Gal,
Constraints from \ensuremath{\Lambda} hypernuclei on the \ensuremath{\Lambda}NN content of the \ensuremath{\Lambda}-nucleus potential,
\href{https://doi:10.1016/j.physletb.2023.137669}{Phys. Lett. B \textbf{837}, 137669 (2023)},
\href{https://arxiv.org/abs/2204.02264}{arXiv:2204.02264~[nucl-th]};
\ensuremath{\Lambda} hypernuclear potentials beyond linear density dependence,
\href{https://doi:10.1016/j.nuclphysa.2023.122725}{Nucl. Phys. A \textbf{1039}, 122725 (2023)},
\href{https://arxiv.org/abs/2306.06973}{arXiv:2306.06973~[nucl-th]}

\bibitem{Hypweb}
  P.~Eckert, P.~Achenbach et al.,
  Chart of hypernuclides -- Hypernuclear structure and decay data (2021),
  \href{https://hypernuclei.kph.uni-mainz.de}{https://hypernuclei.kph.uni-mainz.de}

\bibitem{ParticleDataGroup:2020ssz}
  R.~L.~Workman et al. (Particle Data Group),
  The Reviews of Particle Physics,
  \href{https://doi.org/10.1093/ptep/ptac097}{Prog. Theor. Exp. Phys. 2022, 083C01 (2022)} and 2023 update

\bibitem{STAR:2017gxa}
  L.~Adamczyk et~al.\ (STAR Collaboration),
  Measurement of the ${}_{\Lambda}^{3}\mathrm{H}$ lifetime in Au+Au collisions at the BNL Relativistic Heavy Ion Collider,
  \href{https://doi.org/10.1103/PhysRevC.97.054909}{Phys. Rev. C \textbf{97}, 054909 (2018)},
  \href{https://arxiv.org/abs/1710.00436}{arXiv:1710.00436~[nucl-ex]}

\bibitem{Rappold:2013fic}
  C.~Rappold et~al. (HypHI Collaboration),
  Hypernuclear spectroscopy of products from 6Li projectiles on a carbon target at 2 AGeV,
  \href{https://doi.org/10.1016/j.nuclphysa.2013.05.019}{Nucl. Phys. A \textbf{913}, 170 (2013)},
  \href{https://arxiv.org/abs/1305.4871}{arXiv:1305.4871~[nucl-ex]}

\bibitem{STAR:2021orx}
 M.~Abdallah et~al. (STAR Collaboration),
 Measurements of ${}_{\Lambda}^{3}\mathrm{H}$ and ${}_{\Lambda}^{4}\mathrm{H}$ Lifetimes and Yields in Au+Au Collisions in the High Baryon Density Region,
 \href{https://doi.org/10.1103/PhysRevLett.128.202301}{Phys. Rev. Lett. \textbf{128}, 202301 (2022)},
 \href{https://arxiv.org/abs/2110.09513}{arXiv:2110.09513~[nucl-ex]}

\bibitem{Lock:1964bp}
  \textit{International Conference on Hyperfragments},
  St.~Cergue, Switzerland, 28-30 March, 1963,
  edited by W.~O. Lock,
  \href{https://doi.org/10.5170/CERN-1964-001}{CERN Yellow Reports: Conference Proceedings (CERN, Geneva, 1964)}

\bibitem{PhysRev.180.1307}
  R.~E. Phillips and J.~Schneps,
  Lifetimes of light hyperfragments.\ II,
  \href{https://doi.org/10.1103/PhysRev.180.1307}{Phys. Rev. \textbf{180}, 1307 (1969)}

\bibitem{PhysRevLett.20.819}
  G.~Keyes et al.,
  New measurement of the ${}_{\Lambda}\mathrm{H}^{3}$ lifetime,
  \href{https://doi.org/10.1103/PhysRevLett.20.819}{Phys. Rev. Lett. \textbf{20}, 819 (1968)}

\bibitem{KEYES1970}
  G.~Keyes et al.,
  Properties of ${}_{\Lambda}\mathrm{H}^{3}$,
  \href{https://doi.org/10.1103/PhysRevD.1.66}{Phys. Rev. D \textbf{1}, 66 (1970)}

\bibitem{Bohm:1970se}
  G.~Bohm et~al.,
  On the lifetime of the ${}_{\Lambda}\mathrm{H}^{3}$ hypernucleus,
  \href{https://doi.org/10.1016/0550-3213(70)90335-4}{Nucl. Phys. B \textbf{16}, 46 (1970)},
  [Erratum: Nucl. Phys. B 16, 523 (1970)]

\bibitem{Keyes:1973two}
  G.~Keyes, J.~Sacton, J.~H. Wickens, and M.~M. Block,
  A measurement of the lifetime of the ${}_{\Lambda}\mathrm{H}^{3}$ hypernucleus,
  \href{https://doi.org/10.1016/0550-3213(73)90197-1}{Nucl. Phys. B \textbf{67}, 269 (1973)}

\bibitem{STAR:2010gyg}
  B.~I. Abelev et~al. (STAR Collaboration),
  Observation of an Antimatter Hypernucleus,
  \href{https://doi.org/10.1126/science.1183980}{Science \textbf{328}, 58 (2010)},
  \href{https://arxiv.org/abs/1003.2030}{arXiv:1003.2030~[nucl-ex]}

\bibitem{ALICE:2015oer}
 J.~Adam et~al.\ (ALICE Collaboration),
 ${}^{3}_{\Lambda}\mathrm H$ and ${}^{3}_{\bar{\Lambda}} \overline{\mathrm H}$ production in Pb-Pb collisions at $\sqrt{s_{\rm NN}} =$ 2.76 TeV,
 \href{https://doi.org/10.1016/j.physletb.2016.01.040}{Phys. Lett. B \textbf{754}, 360 (2016)},
 \href{https://arxiv.org/abs/1506.08453}{arXiv:1506.08453~[nucl-ex]}

\bibitem{ALICE:2019vlx}
  S.~Acharya et~al.\ (ALICE Collaboration),
  ${}^3_\Lambda\mathrm{H}$ and ${}^3_{\bar{\Lambda}}\mathrm{\overline{H}}$ lifetime measurement in Pb-Pb collisions at $\sqrt{s_{\mathrm{NN}}} = $ 5.02 TeV via two-body decay,
  \href{https://doi.org/10.1016/j.physletb.2019.134905}{Phys. Lett. B \textbf{797}, 134905 (2019)},
  \href{https://arxiv.org/abs/1907.06906}{arXiv:1907.06906~[nucl-ex]}

\bibitem{Kamada:1997rv}
  H.~Kamada, J.~Golak, K.~Miyagawa, H.~Witala, and W.~Gl{\"o}ckle,
  Pi mesonic decay of the hypertriton,
  \href{https://doi.org/10.1103/PhysRevC.57.1595}{Phys. Rev. C \textbf{57}, 1595 (1998)},
  \href{https://arxiv.org/abs/nucl-th/9709035}{arXiv:9709035~[nucl-th]}

\bibitem{Gal:2018bvq}
  A.~Gal and H.~Garcilazo,
  Towards resolving the ${}_{\Lambda}^{3}\mathrm{H}$ lifetime puzzle,
  \href{https://doi.org/10.1016/j.physletb.2019.02.014}{Phys. Lett. B \textbf{791}, 48 (2019)},
  \href{https://arxiv.org/abs/1811.03842}{arXiv:1811.03842~[nucl-th]}

\bibitem{Hildenbrand:2020kzu}
  F.~Hildenbrand and H.~W. Hammer,
  Lifetime of the hypertriton,
  \href {https://doi.org/10.1103/PhysRevC.102.064002}{Phys. Rev. C \textbf{102}, 064002 (2020)},
  \href{https://arxiv.org/abs/2007.10122}{arXiv:2007.10122~[nucl-th]}

\bibitem{Eckert:2022dyz}
  P.~Eckert et~al.,
  Systematic treatment of hypernuclear data and application to the hypertriton,
  \href{https://doi.org/10.31349/SuplRevMexFis.3.0308069}{Rev. Mex. Fis. Suppl. \textbf{3}, 0308069 (2022)},
  \href{https://arxiv.org/abs/2201.02368}{arXiv:2201.02368~[physics.data-an]}

\bibitem{Rayet:1966fe}
  M.~Rayet and R.~H. Dalitz,
  Lifetime of ${}_{\Lambda}\mathrm{H}^{3}$,
  \href{https://doi.org/10.1007/BF02857527}{Nuovo Cim. A \textbf{46}, 786 (1966)}

\bibitem{Congleton:1992kk}
  J.~G. Congleton,
  A Simple model of the hypertriton,
  \href{https://doi.org/10.1088/0954-3899/18/2/015}{J. Phys. G \textbf{18}, 339 (1992)}

\bibitem{Maessen:1989sx}
 P.~M.~M. Maessen, T.~A. Rijken, and J.~J. de~Swart,
 Soft Core Baryon Baryon One Boson Exchange Models. II. Hyperon - Nucleon Potential,
 \href {https://doi.org/10.1103/PhysRevC.40.2226}{Phys. Rev. C \textbf{40}, 2226 (1989)}

\bibitem{Nogga:2001ef}
 A.~Nogga, H.~Kamada, and W.~Gl{\"o}ckle,
 The Hypernuclei ${}_\Lambda^4\mathrm{He}$ and ${}_\Lambda^4\mathrm{H}$: Challenges for Modern Hyperon Nucleon Forces,
 \href{https://doi.org/10.1103/PhysRevLett.88.172501}{Phys. Rev. Lett. \textbf{88}, 172501 (2002)},
 \href{https://arxiv.org/abs/nucl-th/0112060}{arXiv:0112060~[nucl-th]}

\bibitem{Hildenbrand:2023rkc}
  F.~Hildenbrand,  H.~W.~Hammer,
  Pionic Final State Interactions and the Hypertriton Lifetime,
  \href{https://doi.org/10.1140/epja/s10050-023-01197-z}{Eur. Phys. J. A \textbf{59}, 280 (2024)},
  \href{https://arxiv.org/abs/2309.12822}{arXiv:2309.12822~[nucl-th]}

\bibitem{Golak:1996hj}
  J.~Golak et al.,
  The Non-mesonic weak decay of the hypertriton,
  \href{https://doi.org/10.1103/PhysRevC.56.2892}{Phys. Rev. C \textbf{55}, 2196 (1997)},
  [Erratum: Phys. Rev. C \textbf{56}, 2892 (1997)],
  \href{https://arxiv.org/abs/nucl-th/9612065}{arXiv:9612065~[nucl-th]}

\bibitem{Perez-Obiol:2018oax}
  A.~P\'erez-Obiol, D.~R. Entem, and A.~Nogga,
  $\Lambda N \to NN$ EFT potentials and hypertriton non-mesonic weak decay,
  \href{https://doi.org/10.1088/1742-6596/1024/1/012033}{J. Phys. Conf. Ser. \textbf{1024}, 012033 (2018)}

\bibitem{BESIII:2018cnd}
  M.~Ablikim et~al.\ (BESIII Collaboration),
  Polarization and Entanglement in Baryon-Antibaryon Pair Production in Electron-Positron Annihilation,
  \href{https://doi.org/10.1038/s41567-019-0494-8}{Nature Phys. \textbf{15}, 631 (2019)},
  \href{https://arxiv.org/abs/1808.08917}{arXiv:1808.08917~[hep-ex]}

\bibitem{Donoghue:1992dd}
  J.~F. Donoghue, E.~Golowich, and B.~R. Holstein,
  \href{https://doi.org/10.1017/CBO9780511524370}{\textit{Dynamics of the Standard Model}, Cambridge University Press, Cambridge, 2014, Vol. 2}

\bibitem{Friedman:2007zza}
  E.~Friedman and A.~Gal,
  In-medium nuclear interactions of low-energy hadrons,
  \href {https://doi.org/10.1016/j.physrep.2007.08.002}{Phys. Rept. \textbf{452}, 89 (2007)},
  \href{https://arxiv.org/abs/0705.3965}{arXiv:0705.3965~[nucl-th]}

\bibitem{Friedman:2014msa}
  E.~Friedman and A.~Gal,
  Testing in-medium \ensuremath{\pi N} dynamics on pionic atoms,
  \href{https://doi.org/10.1016/j.nuclphysa.2014.05.017}{Nucl. Phys. A \textbf{928}, 128 (2014)},
  \href{https://arxiv.org/abs/1405.7133} {arXiv:1405.7133~[nucl-th]}

\bibitem{Friedman:2019zhc}
  E.~Friedman and A.~Gal,
  The pion-nucleon \ensuremath{\sigma} term from pionic atoms,
  \href{https://doi.org/10.1016/j.physletb.2019.03.036}{Phys. Lett. B \textbf{792}, 340 (2019)},
  \href{https://arxiv.org/abs/1901.03130}{arXiv:1901.03130~[nucl-th]}

\bibitem{Fri20}
  E.~Friedman, unpublished (2020)

\bibitem{Arndt:2006bf}
  R.~A. Arndt, W.~J. Briscoe, I.~I. Strakovsky, and\ R.~L. Workman,
  Extended partial-wave analysis of \ensuremath{\pi N} scattering data,
  \href{https://doi.org/10.1103/PhysRevC.74.045205}{Phys. Rev. C \textbf{74}, 045205 (2006)},
  \href{https://arxiv.org/abs/nucl-th/0605082}{arXiv:0605082~[nucl-th]}; see also SAID program \href{http://gwdac.phys.gwu.edu/}{http://gwdac.phys.gwu.edu/}

\bibitem{Friedman:2004jh}
  E.~Friedman et~al.,
  The in-medium isovector \ensuremath{\pi N} amplitude from low energy pion scattering,
  \href{https://doi.org/10.1103/PhysRevLett.93.122302}{Phys. Rev. Lett. \textbf{93}, 122302 (2004)},
  \href{https://arxiv.org/abs/nucl-ex/0404031}{arXiv:0404031~[nucl-ex]}

\bibitem{Friedman:2005pt}
  E.~Friedman et~al.,
  Elastic scattering of low energy pions by nuclei and the in-medium isovector \ensuremath{\pi N} amplitude,
  \href{https://doi.org/10.1103/PhysRevC.72.034609}{Phys. Rev. C \textbf{72}, 034609 (2005)},
  \href{https://arxiv.org/abs/nucl-ex/0507008}{arXiv:0507008~[nucl-ex]}

\bibitem{Barrett2013}
  B.~R. Barrett, P.~Navrátil, and J.~P. Vary,
  Ab initio no core shell model,
  \href{https://doi.org/10.1016/j.ppnp.2012.10.003}{Prog. Part. Nucl. Phys. \textbf{69}, 131 (2013)}

\bibitem{Wirth:2017bpw}
  R.~Wirth, D.~Gazda, P.~Navrátil, and\ R.~Roth,
  Hypernuclear No-Core Shell Model,
  \href{https://doi.org/10.1103/PhysRevC.97.064315}{Phys. Rev. \textbf{C97}, 064315 (2018)},
  \href{https://arxiv.org/abs/1712.05694}{arXiv:1712.05694~[nucl-th]}

\bibitem{Navratil:1999pw}
  P.~Navr\'atil, G.~P. Kamuntavicius, and B.~R.\ Barrett,
  Few nucleon systems in translationally invariant harmonic oscillator basis,
  \href{https://doi.org/10.1103/PhysRevC.61.044001}{Phys. Rev. \textbf{C61}, 044001 (2000)},
  \href{https://arxiv.org/abs/nucl-th/9907054}{arXiv:9907054~[nucl-th]}

\bibitem{Carlsson:2015vda}
  B.~D. Carlsson et al.,
  Uncertainty analysis and order-by-order optimization of chiral nuclear interactions,
  \href {https://doi.org/10.1103/PhysRevX.6.011019}{Phys. Rev. \textbf{X6}, 011019 (2016)},
  \href{https://arxiv.org/abs/1506.02466}{arXiv:1506.02466~[nucl-th]}

\bibitem{Polinder:2006zh}
  H.~Polinder, J.~Haidenbauer, and U.-G. Mei{\ss}ner, Hyperon-nucleon interactions: A Chiral effective field theory approach,
  \href{https://doi.org/10.1016/j.nuclphysa.2006.09.006}{Nucl. Phys. \textbf{A779}, 244 (2006)},
  \href{https://arxiv.org/abs/nucl-th/0605050}{arXiv:0605050~[nucl-th]}

\bibitem{Haidenbauer:2007ra}
  J.~Haidenbauer, U.-G. Mei{\ss}ner, A.~Nogga, and H.~Polinder,
  The Hyperon-nucleon interaction: Conventional versus effective field theory approach,
  in \textit{Topics in Strangeness Nuclear Physics},
  \href{https://doi.org/10.1007/978-3-540-72039-3_4}{Lecture Notes in Physics Vol.~\textbf{724} (Springer, Berlin, 2007), p.~113},
  \href{https://arxiv.org/abs/nucl-th/0702015}{arXiv:0702015~[nucl-th]}

\bibitem{Htun:2021jnu}
  T.~Y.~Htun, D.~Gazda, C.~Forss\'en, and\ Y.~Yan,
  Systematic Nuclear Uncertainties in the Hypertriton System,
  \href {https://doi.org/10.1007/s00601-021-01675-4}{Few Body Syst. \textbf{62}, 94 (2021)},
  \href{https://arxiv.org/abs/2109.09479}{arXiv:2109.09479~[nucl-th]}

\bibitem{Gazda:2022fte}
  D.~Gazda, T.~Y.~Htun, and C.~Forss\'en,
  Nuclear physics uncertainties in light hypernuclei,
  \href {https://doi.org/10.1103/PhysRevC.106.054001}{Phys. Rev. C \textbf{106}, 054001 (2022)},
  \href{https://arxiv.org/abs/2208.02176}{arXiv:2208.02176~[nucl-th]}

\bibitem{Wendt2015}
  K.~A. Wendt, C.~Forssén, T.~Papenbrock, and D.~Sääf,
  Infrared length scale and extrapolations for the no-core shell model,
  \href{https://doi.org/10.1103/PhysRevC.91.061301}{Phys. Rev. C \textbf{91}, 061301(R) (2015)},
  \href{https://arxiv.org/abs/1503.07144}{arXiv:1503.07144~[nucl-th]}

\bibitem{Forssen:2017wei}
  C.~Forssén et al.,
  Large-scale exact diagonalizations reveal low-momentum scales of nuclei,
  \href{https://doi.org/10.1103/PhysRevC.97.034328}{Phys. Rev. C \textbf{97}, 034328 (2018)},
  \href{https://arxiv.org/abs/1712.09951}{arXiv:1712.09951~[nucl-th]}

\bibitem{Konig:2014hma}
  S.~K{\"o}nig et al.,
  Ultraviolet extrapolations in finite oscillator bases,
  \href{https://doi:10.1103/PhysRevC.90.064007}{Phys. Rev. C \textbf{90}, 064007 (2014)},
  \href{https://arxiv.org/abs/1409.5997}{arXiv:1409.5997~[nucl-th]}

\bibitem{STAR:2019wjm}
  J.~Adam et~al.\ (STAR Collaboration),
  Measurement of the mass difference and the binding energy of the hypertriton and antihypertriton,
  \href{https://doi.org/10.1038/s41567-020-0799-7}{Nature Phys. \textbf{16}, 409 (2020)},
  \href{https://arxiv.org/abs/1904.10520}{arXiv:1904.10520~[hep-ex]}

\end{thebibliography}
\end{document}